\documentclass[aps,prl,twocolumn,superscriptaddress,showpacs,amsmath,amssymb,longbibliography]{revtex4-1}
\usepackage{graphicx}
\usepackage[breaklinks,colorlinks,linkcolor=blue,citecolor=blue,urlcolor=blue]{hyperref}
\usepackage[english]{babel}
\usepackage{amsmath}
\usepackage{xcolor}
\usepackage{ulem}

\usepackage{color}
\definecolor{nred} {RGB}{224,0,0}
\definecolor{nblue} {RGB}{28,130,185}
\definecolor{dgreen} {RGB}{28,150,50}
\definecolor{npurple} {RGB}{125, 35, 251}

\usepackage[sort&compress]{natbib}
\begin{document}

\title{Resistivity and its fluctuations in disordered many-body systems:\\from chains to planes} 

\author{M. Mierzejewski}
\affiliation{Department of Theoretical Physics, Faculty of Fundamental Problems of Technology, Wroc\l aw University of Science and Technology, 50-370 Wroc\l aw, Poland}

\author{M. {\'S}roda}
\affiliation{Department of Theoretical Physics, Faculty of Fundamental Problems of Technology, Wroc\l aw University of Science and Technology, 50-370 Wroc\l aw, Poland}

\author{J. Herbrych}
\affiliation{Department of Theoretical Physics, Faculty of Fundamental Problems of Technology, Wroc\l aw University of Science and Technology, 50-370 Wroc\l aw, Poland}

\author{P. Prelov\v sek}
\affiliation{J. Stefan Institute, SI-1000 Ljubljana, Slovenia }
\affiliation{Faculty of Mathematics and Physics, University of Ljubljana, SI-1000 Ljubljana, Slovenia }

\begin{abstract}
We study a quantum particle coupled to hard-core bosons and  propagating on disordered ladders with $R$ legs. The particle dynamics is studied with the help of rate equations for the boson-assisted transitions between the Anderson states.  We demonstrate that for finite $R < \infty$ and sufficiently strong disorder the dynamics is subdiffusive, while the two-dimensional planar systems with $R\to \infty$ appear to be diffusive for arbitrarily strong disorder. The transition from diffusive to subdiffusive regimes may be identified via statistical fluctuations of resistivity. The corresponding distribution function in the diffusive regime has fat tails which decrease with the system size $L$ much slower than $1/\sqrt{L}$. Finally, we present evidence that similar non--Gaussian fluctuations  arise also in standard models of many-body localization, i.e., in strongly disordered quantum spin chains.
\end{abstract}

\date{\today}

\maketitle

{\it Introduction--}
There is a vast numerical evidence supporting 
the presence of many-body localization (MBL) \cite{basko06,oganesyan07} in strongly disordered one-dimensional systems (1D), such as spin chains or equivalent models of interacting spinless fermions \cite{monthus10,luitz15,ZZZ5_4,Ponte2015,lazarides15,vasseur15a,serbyn2014a,pekker2014,torres15,torres16,laumann2015,huse14,gopal17,Hauschild_2016,herbrych13,imbrie16,steinigeweg16,Herbrych17}. Furthermore, disorder-induced localization is consistent with several experimental studies of cold-atoms in optical lattices  \cite{kondov15,schreiber15,choi16,bordia16,bordia2017_1,smith2016}. Strongly disordered systems  exhibit very slow relaxation \cite{znidaric08,bardarson12,kjall14,serbyn15,luitz16,serbyn13_1,bera15,altman15,agarwal15,gopal15,znidaric16,mierzejewski2016,lev14,lev15,barisic16,bonca17,bordia2017_1,zakrzewski16,protopopov2018,sankar2018,zakrzewski2018} that shows up also in systems which are not localized, e.g., due to too weak disorder or due to the SU(2) spin--symmetry \cite{Chandran2014,potter16,prelovsek16,proto2017,friedman2017}. Then, the dynamics is typically subdiffusive \cite{luitz2016prl,luitz116,znidaric16,gopal17,kozarzewski18,prelovsek217,new_karrasch,prelovsek2018a}, what is frequently considered as a precursor to localization \cite{luitz2016prl,luitz116,znidaric16,gopal17,kozarzewski18,prelovsek217,new_karrasch,prelovsek2018a} and can be attributed to the presence of the so-called weak links \cite{agarwal15,bordia2017_1,agarwal16,luschen17}.

The transport properties of strongly disordered systems in higher dimensions are by far less explored. On the one hand, results in Ref.~\cite{derek} suggest that MBL is stable only in 1D systems provided that interactions decay exponentially with distance. On the other hand, the experiments show signatures of localization also in two-dimensional (2D) \cite{choi16,bordia2017_1} and three-dimensional \cite{kondov15} systems. Thus, the dynamics of strongly disordered systems beyond 1D remains largely an open problem. Here, the main challenge is that most numerical methods allow the study of too small systems or too short evolution times to judge on the long-time properties of macroscopic systems.
 
\begin{figure}
\includegraphics[width=\columnwidth]{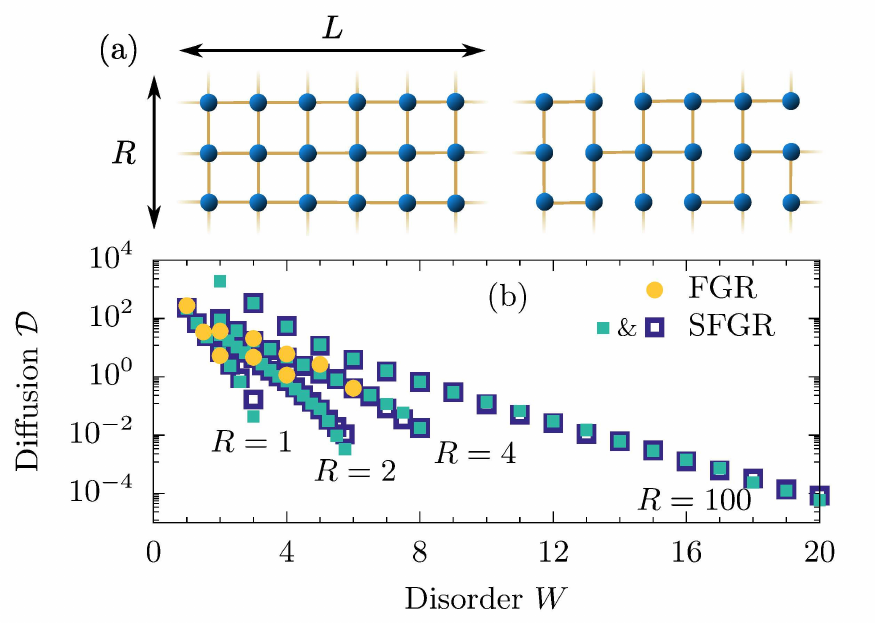}
\caption{(a) Sketch of the system (left) and percolation problem (right). (b) Diffusion constant ${\cal D}$ obtained from rate equations with Fermi golden rule (FGR) for $L=200$ (circles) and simplified-FGR (SFGR) for $N=LR=10^4$ (squares) with various number of legs, $R$, up to a 2D system for $R \gg 1$. Open and filled squares correspond to different realizations of disorder.}
\label{fig1}
\end{figure}

In order to approach the MBL physics beyond 1D, we study a simpler many-body system, i.e., a single quantum particle coupled to hard-core bosons. The particle propagates on a disordered $R$-leg ladder with different number of legs, ranging from $R=1$ (chains) to $R\to \infty$ (planes), see Fig.~\ref{fig1}(a). The system's dynamics is modeled via rate equations emerging from the Fermi golden rule (FGR) for transitions between the localized Anderson states \cite{prelovsek2018a,Mierzejewski2019}. The approach is simple enough that we are able to obtain unbiased numerical results for rather large systems with $N \sim10^4$ sites and up to $R =10^2$ legs. Previous studies of the same Hamiltonian on a single chain ($R=1$) revealed that for strong disorder the particle dynamics is subdiffusive \cite{bonca2018} and that such dynamics may be well described within the FGR approach \cite{prelovsek2018a,Mierzejewski2019}. 

Here,  we show that sufficiently strong disorder causes a transition between diffusive and subdiffusive regimes for arbitrary $R < \infty$. For weaker disorder, the diffusion constant $\cal{D}$ decreases almost exponentially with increasing disorder and is a self-averaging quantity with respect to various realizations of the disorder. Namely, the sample-to-sample fluctuations  of $\cal{D}$ are Gaussian and its width decreases with system length $L$ as $1/\sqrt{L}$. However, at the transition to subdiffusion we observe strong non--Gaussian fluctuations of effective resistivity, defined here as the inverse diffusion constant, $\rho={\cal D}^{-1}$. As a consequence, the probability distribution of $\rho$ reveals fat tails, $f(\rho)\propto \rho^{-2}$, with size dependence much weaker than $1/\sqrt{L}$. In order to test whether such statistical fluctuations arise only in the studied model, possibly as an artifact of the FGR, we numerically calculate the distributions of ${\cal D} $ within prototype quantum models of MBL. Our results suggest that the fat-tailed statistical fluctuations of resistivity are generic for strongly disordered quasi-1D quantum models. 
 
{\it Particle in a disordered potential coupled to hard-core bosons--} 
We study a quantum particle on a ladder containing $R$ legs of length $L$ coupled to itinerant hard-core bosons. The system is described by the Hamiltonian \cite{Mierzejewski2019},
\begin{eqnarray}
H &=& - t \sum_{\langle i,j \rangle} c^\dagger_{i} c^{\phantom{\dagger}}_j + \sum_j \varepsilon_j n_j + g \sum_j n_j ( a^\dagger_j + a^{\phantom{\dagger}}_j ) \nonumber \\
& & +\omega_0 \sum_j a_j^\dagger a^{\phantom{\dagger}}_j - t_b \sum_{\langle i,j \rangle} a_{i}^\dagger a^{\phantom{\dagger}}_j,
\label{ham} 
\end{eqnarray} 
where $\varepsilon_j$ are independent random potentials uniformly distributed in $[-W,W]$. Here, $c^\dagger_{j}$ and $a_{j}^\dagger$ refer to local fermion and hard-core boson operators ($a^\dagger_j a^\dagger_j =0$), respectively.  For simplicity, we set $t=1$, $\omega_0=g=1$, $t_b=0.2$ and restrict our studies to the case of an infinite temperature, $\beta=1/k_{\rm B}T \to 0$.

In order to derive the rate equations (RE), we first diagonalize the single-particle part of the Anderson Hamiltonian [first two terms in Eq.~(\ref{ham})], $H_{\rm SP}=\sum_l \epsilon_l \varphi_l^\dagger \varphi_l^{\phantom{\dagger}} $, where $ \varphi_l = \sum_i \phi_{li} c_i$ and $ \phi_{li}$ are single-particle eigen-functions. We then use the FGR to calculate the transition rates $\Gamma_{lk}$ between different $l\neq k$ Anderson states $|l\rangle=\varphi_l^\dagger |0\rangle$. The emerging RE allow us to study large system sizes $N=LR \lessapprox 10^3$, whereas for $N \sim 10^4$ we use a simplified FGR (SFGR). In the latter approach, we neglect the momentum dependence of matrix elements for particle-boson interaction and assume a uniform bosonic density of states. In the case of a single chain, the explicit form of $\Gamma_{lk}$ has been derived in \cite{prelovsek2018a} and \cite{Mierzejewski2019} for FGR and SFGR, respectively. For convenience, we recall the main steps of derivations in the Supplemental Material~\cite{suppmat}.

To directly address the transport, we consider an open system introducing the current source at the left rung and the current drain at the right rung of the ladder, i.e., we study a system with current flowing (on average) along the legs, as described by the RE,
\begin{equation}
\frac{{\rm d} n_l}{{\rm d} t} = I_l+\sum_{k\ne l} ( \Gamma_{kl} n_{k} - \Gamma_{lk} n_l). \label{req} 
\end{equation} 
Here, $n_l$ is the occupation of the state $|l\rangle$ and $I_l=I^{s}_l+I^{d}_l$ accounts for the source and the drain, respectively,
\begin{equation}
I^s_l= {\cal I}_0\sum_{i \in \rm left} |\phi_{li}|^2,\quad \quad I^s_d=- {\cal I}_0 \sum_{i \in \rm right} |\phi_{li}|^2,
\label{isd}
\end{equation}
where the summations are carried out over the left- and right-edge rungs. Since $\phi_{li}$ are normalized, the total injected current $\sum_l I^s_l=R\,{\cal I}_0$ hence, ${\cal I}_0$ is the average current density. Then, the diffusion constant ${\cal D}$ is obtained from the relation between the current density and the gradient of the particle density, ${\cal D}=- {\cal I}_0/ \nabla n_i$ with $n_i=\sum_l n_l |\phi_{li}|^2$ and
$n_l$ representing the stationary solution of RE (\ref{req}). We refer to the Supplemental Material~\cite{suppmat} for technical details on the stationary solution. 

Fig.~\ref{fig1}(b) shows ${\cal D}$ vs disorder $W$. Each data set corresponds to a single realization of disorder and varying $W$. We also compare results obtained from FGR for $L=200$ and SFGR with much larger $L$. Few things become apparent from the presented results:  finite-size effects and sample-to-sample fluctuations are negligible in the diffusive regime; the simplifications introduced within SFGR  do not influence the qualitative results. Another evident but nontrivial result is the exponential dependence of ${\cal D}$ on the disorder strength $W$ in a wide range of the latter \cite{barisic10, prelovsek116,lev2019}, apparently extending to very large $W$ for the 2D system, $R\gg1$.

\begin{figure} 
\includegraphics[width=\columnwidth]{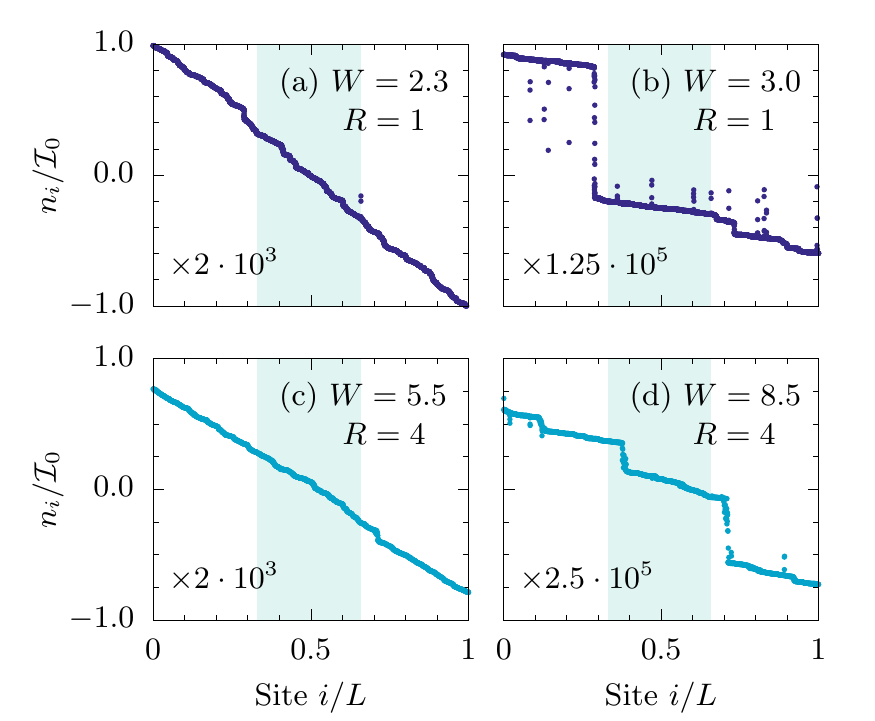}
\caption{Spatial profiles of $n_i$ for $N=10^4$. Panels (a), (b) and (c), (d)  show results for $R=1$ and $R=4$, respectively. Shaded regions represent sections of the system where the diffusion constant is calculated. Results for $R > 1$ are averaged over the rungs.}
\label{fig2}
\end{figure} 

 Results in Fig. \ref{fig1}b are restricted  to sufficiently weak disorder when  the spatial variation of $n_i$ along the legs is linear, as shown in Figs.~\ref{fig2}(a) and \ref{fig2}(c). However, for stronger disorder $W \sim W_c$, the variation becomes nonlinear due to the formation of weak links which are clearly exemplified in Figs.~\ref{fig2}(b) and \ref{fig2}(d). Such behavior signals a transition between the diffusive and subdiffusive regimes. The threshold value $W_c$ increases with $R$, but apparently remains finite provided that $R< \infty$. Fig.~\ref{fig1}(b) shows also that differences between results for various realizations of disorder \cite{barisic16,Husex2017} increase upon approaching the transition.  Next we demonstrate that the latter sample-to-sample fluctuations are universal for the transition between the diffusive and the subdiffusive regimes. 

{\it Sample-to-sample fluctuations--} 
In order to explain the statistical fluctuations of ${\cal D}$, we first consider a strongly disordered single chain ($R=1$) where,  for simplicity,  the FGR transitions are restricted to Anderson states on neighboring sites, $\Gamma_{kl} \sim \delta_{k,l+1} $, $I^s_l \simeq {\cal I}_0 \delta_{l1}$ and $I^d_l \simeq {\cal I}_0 \delta_{lL}$. Then, one derives from the stationary solution of  Eq.~(\ref{req}) that $n_{l}-n_{l+1}={\cal I}_0/\Gamma_{l,l+1}$, and consequently
\begin{equation}
\rho={\cal D}^{-1} \simeq \frac{ n_1-n_L}{L {\cal I}_0}=\frac{1}{L}\sum_{l}\tau_l,\quad \quad \tau_l=\frac{1}{\Gamma_{l,l+1}}. \label{ddef}
\end{equation}
As previously demonstrated  for the same model \cite{prelovsek2018a,Mierzejewski2019}, the transition times $\tau_l=\Gamma^{-1}_{l,l+1}$, can be well approximated via independent random variables with power-law probability distribution function $f_{\tau}(\tau)\propto \tau^{-(\alpha+1)}$ for large enough $\tau$. Within this simplification, $\rho$ in Eq.~(\ref{ddef}) becomes an average of $L$-independent random variables with the transition to subdiffusion at $\alpha=1$. 
 
Here, we focus on the diffusive regime, $1<\alpha <2$, where the average transition time is finite \mbox{$\langle \tau \rangle = \int_0^{\infty} {\rm d} \tau \; f_{\tau}(\tau) \tau< \infty$}, but $\langle \tau^2 \rangle$ diverges, thus the fluctuations of $\rho$ are non--Gaussian. It is well established for the fat-tailed (the so-called $\alpha$-stable) distributions \cite{bouchaud89} that the random variable 
\begin{equation}
u=\frac{1}{L^{1/\alpha}} \left(\sum_{l=1}^L \tau_l-L \langle \tau \rangle \right)= L^{(\alpha-1)/\alpha} (\rho - \langle \tau \rangle), 
\label{udef}
\end{equation}
has a limit distribution $f_u(u)$ for $L\rightarrow \infty $ and asymptotically $f_u(u) \propto u^{-(\alpha+1)} $. Clearly, the latter determines the tails as well as the $L$-dependence of the resistivity distribution $f_{\rho}( \rho)$. In particular, close to the transition to the subdiffusive regime, $\alpha \rightarrow 1$, the exponent  in r.h.s. of Eq.~(\ref{udef}) vanishes, $(\alpha-1)/\alpha \rightarrow 0$. As a consequence, one obtains weak, at most logarithmic, $L$-dependence of $f_{\rho}(\rho)$. The fat tails can be observed from the cumulative and the complementary cumulative distribution functions of ${\cal D}$ and $\rho$, respectively,
\begin{eqnarray}
F_{\cal D}({\cal D}) & = & \int_0^{{\cal D}} {\rm d} {\cal D}' f_{{\cal D}'}( {\cal D}') \simeq \frac{ {\cal D}^{\alpha}}{\alpha L^{\alpha-1}}, 
\quad {\cal D} \ll \langle \tau \rangle^{-1}, \label{cdf} \quad \\
F^{c}_{\rho}(\rho) & = &\int_{\rho}^{\infty} {\rm d} \rho' f_{\rho}( \rho') \simeq \frac{1}{\alpha L^{\alpha-1}\rho ^{\alpha}}, \quad \rho \gg \langle \tau \rangle\,. \label{ccdf}
\end{eqnarray} 

\begin{figure} 
\includegraphics[width=\columnwidth]{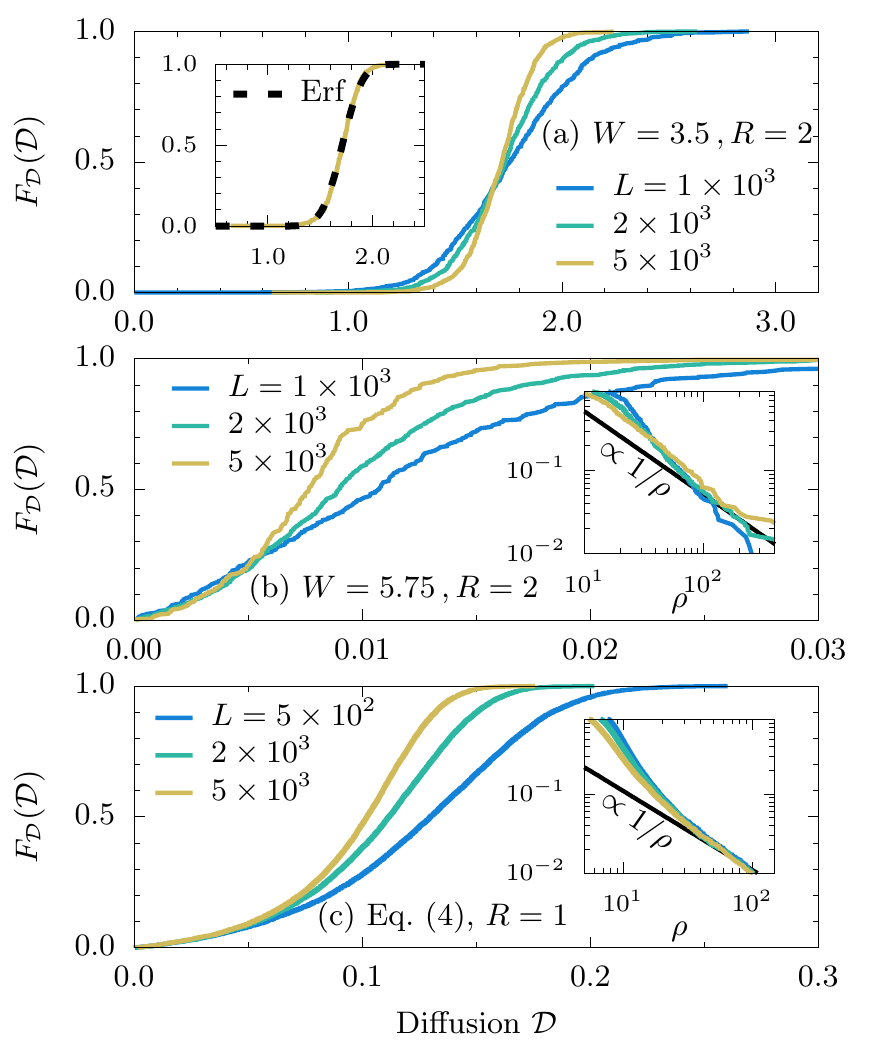}
\caption{Main panels: cumulative distribution functions of the diffusion constant, $F_{\cal D}({\cal D})$. Inset (a):  $F_{\cal D}({\cal D})$ for $L=5\times 10^3$ fitted with the error function. Insets (b),(c): complementary distribution functions of the resistivity $F^c_{\rho}(\rho)$.  (a) and (b) Results obtained via rate equations with SFGR for $R=2$. (c) Results for a chain  with independent random transition times with fat-tailed distribution, Eq.~(\ref{ddef}).}
\label{fig3}
\end{figure} 

It is by far not clear whether the same properties survive in the considered system, when the transition rates are not independent random variables connecting only neighboring sites but, instead, are obtained fully from FGR. In figures \ref{fig3}a and \ref{fig3}b we present $F_{\cal D}({\cal D})$ (main panels) and $F^c_{\rho}(\rho)$ (inset in b) calculated for a two-leg ladder ($R=2$) directly from the stationary solution of Eqs.~(\ref{req}) and with SFGR transition rates. For comparison, we display in Fig.~\ref{fig3}c similar results, obtained from the toy model, Eq.~(\ref{ddef}), with $f_{\tau}(\tau) = \tau^{-(\alpha+1)}$ for $\tau \ge 1$, where we used $\alpha=1.01$. For modest disorder shown in Fig.~\ref{fig3}a we confirm that $F_{\cal D}({\cal D})$ represents an error function, in agreement with the Gaussian fluctuations of ${\cal D}$ and its width decreasing approximately as $1/\sqrt{L}$ (not shown). However, upon approaching the transition to the subdiffusive regime, as in Fig.~\ref{fig3}b, $F_{\cal D}({\cal D})$ clearly differs from the Gaussian case. Results for $F^c_{\rho}(\rho)$ and $F_{\cal D}({\cal D})$ now agree with the analytical predictions, Eqs.~(\ref{ccdf}) and Eqs.~(\ref{cdf}) for $\alpha \rightarrow 1$. In particular, the statistical fluctuations for $\rho \gg \langle \tau \rangle$ (or ${\cal D} \to 0$) only weakly depend on $L$. Moreover, the latter results are qualitatively similar to those shown in Fig.~\ref{fig3}(c) for the toy model ($R=1$) with random transition rates between neighboring Anderson states.

{\it Diffusivity of the planar system--} The  toy model also offers a simple explanation of why the  2D system remains diffusive for arbitrary $W$, as shown in Fig.~\ref{fig1}b. To this end, we construct the lower bound on ${\cal D}$ and  demonstrate that it is non-zero. We consider a network with only nearest-neighbor transitions, shown in the right panel of Fig.~\ref{fig1}a. We set a threshold transition time $\tau_{\rm th} < \infty$ and check $\tau_l$ on each link in the network. For links with $\tau_l < \tau_{\rm th}$ we replace $\tau_l$ with $\tau_{\rm th}$ and remove links which do not satisfy the latter inequality.  As a consequence, the values of all $\tau_l$ increase, hence we end up with a percolation problem for a system which obviously has smaller ${\cal D}$ than the original system. The density of removed links $\int_{\tau_{\rm th}}^{\infty} {\rm d} \tau \; f_{\tau}(\tau)=1/(\alpha \tau_{\rm th} ^{\alpha})$ may be tuned to an arbitrarily small number via increasing $\tau_{\rm th}$, thus the system may be tuned above the percolation threshold for arbitrary $\alpha >0$. Consequently, the transport is always diffusive.
 
{\it Fluctuations in disordered spin chains--} Next, we check whether such anomalous fluctuations are general and arise also beyond the semi-classical RE approach.  To this end, we investigate the sample-to-sample fluctuation of the transport quantities in prototype 1D models which, for strong enough disorder,  exhibit MBL or a diffusion-subdiffusion transition. 

As a first example, we consider  the {\it standard} model of MBL, i.e., the Heisenberg model with quenched disorder introduced via a random on-site magnetic field \cite{basko06,oganesyan07}. It is commonly accepted  that the transition from ergodic to non-ergodic phase takes place at $W/J\simeq 3.7$, where $J$ is the antiferromagnetic exchange coupling  \cite{luitz15}. Furthermore, it has been argued that the MBL phase in this model is preceded by the subdiffusive Griffiths phase \cite{agarwal15,bordia2017_1,agarwal16,luschen17}. The second investigated model describes  the spin dynamics in the Hubbard chain with a random  charge potential. The latter  disorder localizes the charge (i.e., the density of fermions), yielding only the spin degrees of freedom mobile. The effective model  \cite{kozarzewski18,sroda19,kozarzewski2019,protopopov2019} takes a form of the random-exchange ferromagnetic Heisenberg model with a singular distribution of $J$, $f_J(J)=\lambda J^{\lambda-1}$ for $0\leqslant J \leqslant 1$. It was shown that for  strong charge disorder  ($\lambda < 1$)  the spin dynamics   in this specific random-$J$ Heisenberg chain  is subdiffusive \cite{kozarzewski18,sroda19,kozarzewski2019}. Finally, we examine  also the energy transport in the random-transverse-field Ising model for which the existence of MBL has been shown analytically \cite{imbrie14,imbrie16}. 

In order to extract the analogues of the diffusion constant, we  calculate the low-frequency regular part of the conductivity ${\cal D}=C(\omega\to 0)$, where $C$ is the spin conductivity in the Heisenberg models and the thermal conductivity in the transverse-field Ising model. It is important to note that the spectrum of a finite-size system with discrete Hilbert space has to be artificially broadened in order to address the d.c.  limit. Such broadening can, in principle, affect the value of ${\cal D}$. Our results indicate that although the median,  ${\cal D}_{\rm med}$,  may substantially dependent on the broadening, the functional form of the distribution $f_{\cal D}({\cal D})$ does not. We refer to the Supplemental Material \cite{suppmat}  and  Ref. \cite{prelovsek13} for technical details.

\begin{figure} 
 \includegraphics[width=\columnwidth]{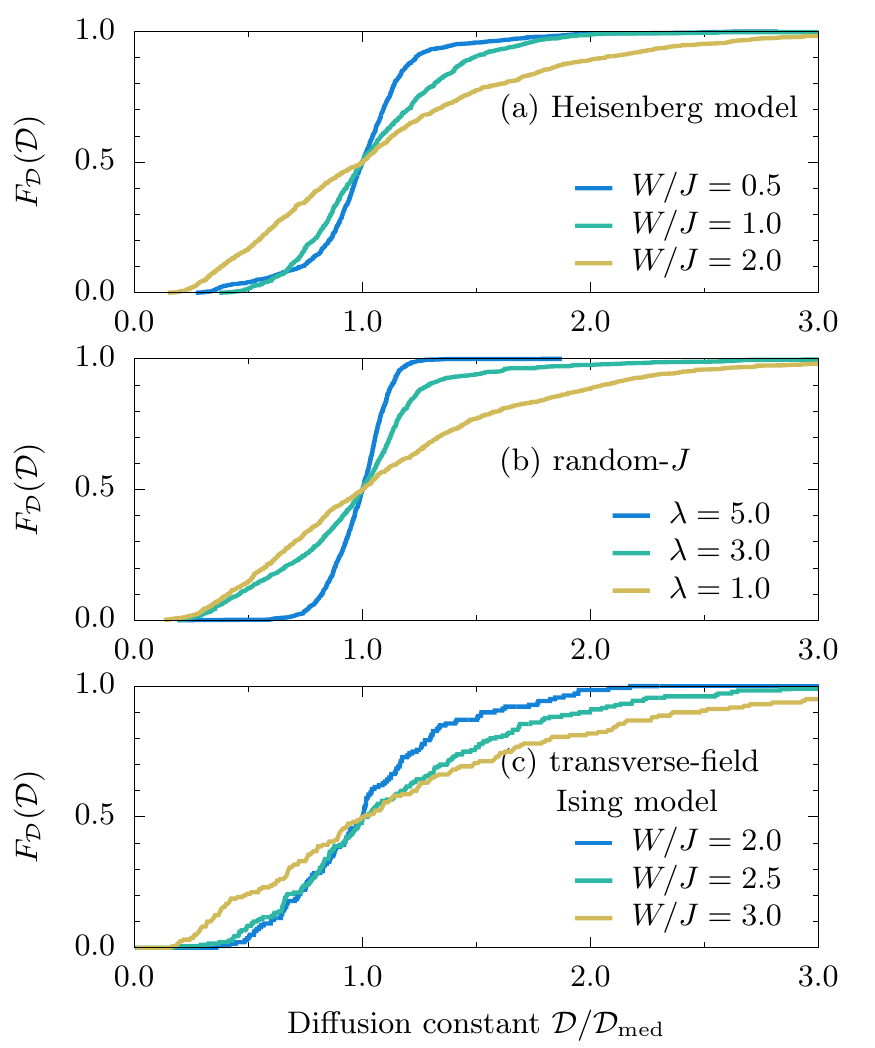}
\caption{Cumulative distribution functions $F_{\cal D}({\cal D})$  as calculated for various disorder strengths in the ergodic regime for (a) random-field Heisenberg model, (b)  random-$J$ Heisenberg model with  $f_J(J)=\lambda J^{\lambda-1}$ distribution and (c) random-transverse-field Ising model (see the text for details). Panels (a) and (b) are obtained with $N_r=1000$ realizations of the disorder, while $N_r=200$ in panel (c).} 
\label{fig4}
\end{figure} 

In Fig.~\ref{fig4}, we present the cumulative distribution functions  $F_{\cal D}({\cal D})$ obtained for the disordered quantum  spin chains. For small enough disorder and for all considered models, $F_{\cal D}({\cal D})$ may be well fitted by the error function, reflecting the Gaussian distribution of ${\cal D}$. On the other hand, increasing the disorder strength changes the functional form of $f_{\cal D}({\cal D})$. It is evident  that the distributions become non-Gaussian, closely resembling the results in Figs. \ref{fig3}b and  \ref{fig3}c for the RE approach. The latter similarity  suggests that  the fat-tailed fluctuations of resistivity at the diffusion-subdiffusion transition are generic for strongly disordered quasi-1D systems. Due to the limitations of the numerical method, we do not get irrefutable evidence for the spin chains and the latter claim should be considered as a well justified conjecture. The numerical verification of  the weak $L$-dependence of $f_{\cal D}({\cal D})$  seems to be a particularly challenging problem.  

{\it Conclusions--}  
We have studied how the transport properties of a strongly disordered system with many-body interactions depend on its dimensionality. The geometry of the $R$-leg ladders allowed tuning the system between one-dimensional ($R=1$) and two-dimensional ($R \to \infty$) cases. On the one hand, we have demonstrated that sufficiently strong disorder causes subdiffusive transport  for any finite $R$ and that the weak-link scenario survives also for $R>1$. On the other hand, planar systems  ($R \to \infty$) appear to be always diffusive, albeit the diffusion constant decreases exponentially with the disorder strength and may eventually become undetectably small. We have shown that the diffusion-subdiffusion transition may be identified via fat-tailed statistical fluctuations of resistivity between different realizations of disorder.  Numerical results obtained for various models of disordered spin chains suggest that the latter fluctuations may be generic for quasi-1D quantum systems.  The presence of non-Gaussian and almost size--independent fluctuations poses challenging problem for numerical studies, especially when self-averaging quantities are obtained numerically from averaging over various realizations of disorder.

\acknowledgments 
M.M. and M.{\'S}. acknowledge support from the National Science Centre, Poland via project 2016/23/B/ST3/00647. In addition, M.{\'S}. and J.H. acknowledge grant support by the Polish National Agency of Academic Exchange (NAWA) under contract PPN/PPO/2018/1/00035. P.P. acknowledges the project N1-0088 of the Slovenian Research Agency. Calculations have been carried out using resources provided by Wroclaw Centre for Networking and Supercomputing.

\bibliography{ref_mbl}

\begin{thebibliography}{73}%
\makeatletter
\providecommand \@ifxundefined [1]{%
 \@ifx{#1\undefined}
}%
\providecommand \@ifnum [1]{%
 \ifnum #1\expandafter \@firstoftwo
 \else \expandafter \@secondoftwo
 \fi
}%
\providecommand \@ifx [1]{%
 \ifx #1\expandafter \@firstoftwo
 \else \expandafter \@secondoftwo
 \fi
}%
\providecommand \natexlab [1]{#1}%
\providecommand \enquote  [1]{``#1''}%
\providecommand \bibnamefont  [1]{#1}%
\providecommand \bibfnamefont [1]{#1}%
\providecommand \citenamefont [1]{#1}%
\providecommand \href@noop [0]{\@secondoftwo}%
\providecommand \href [0]{\begingroup \@sanitize@url \@href}%
\providecommand \@href[1]{\@@startlink{#1}\@@href}%
\providecommand \@@href[1]{\endgroup#1\@@endlink}%
\providecommand \@sanitize@url [0]{\catcode `\\12\catcode `\$12\catcode
  `\&12\catcode `\#12\catcode `\^12\catcode `\_12\catcode `\%12\relax}%
\providecommand \@@startlink[1]{}%
\providecommand \@@endlink[0]{}%
\providecommand \url  [0]{\begingroup\@sanitize@url \@url }%
\providecommand \@url [1]{\endgroup\@href {#1}{\urlprefix }}%
\providecommand \urlprefix  [0]{URL }%
\providecommand \Eprint [0]{\href }%
\providecommand \doibase [0]{http://dx.doi.org/}%
\providecommand \selectlanguage [0]{\@gobble}%
\providecommand \bibinfo  [0]{\@secondoftwo}%
\providecommand \bibfield  [0]{\@secondoftwo}%
\providecommand \translation [1]{[#1]}%
\providecommand \BibitemOpen [0]{}%
\providecommand \bibitemStop [0]{}%
\providecommand \bibitemNoStop [0]{.\EOS\space}%
\providecommand \EOS [0]{\spacefactor3000\relax}%
\providecommand \BibitemShut  [1]{\csname bibitem#1\endcsname}%
\let\auto@bib@innerbib\@empty
\bibitem [{\citenamefont {Basko}\ \emph {et~al.}(2006)\citenamefont {Basko},
  \citenamefont {Aleiner},\ and\ \citenamefont {Altshuler}}]{basko06}%
  \BibitemOpen
  \bibfield  {author} {\bibinfo {author} {\bibfnamefont {D.M.}\ \bibnamefont
  {Basko}}, \bibinfo {author} {\bibfnamefont {I.L.}\ \bibnamefont {Aleiner}}, \
  and\ \bibinfo {author} {\bibfnamefont {B.L.}\ \bibnamefont {Altshuler}},\
  }\bibfield  {title} {\enquote {\bibinfo {title} {Metal--insulator transition
  in a weakly interacting many-electron system with localized single-particle
  states},}\ }\href {\doibase 10.1016/j.aop.2005.11.014} {\bibfield  {journal}
  {\bibinfo  {journal} {Ann. Phys.}\ }\textbf {\bibinfo {volume} {321}},\
  \bibinfo {pages} {1126--1205} (\bibinfo {year} {2006})}\BibitemShut {NoStop}%
\bibitem [{\citenamefont {Oganesyan}\ and\ \citenamefont
  {Huse}(2007)}]{oganesyan07}%
  \BibitemOpen
  \bibfield  {author} {\bibinfo {author} {\bibfnamefont {V.}~\bibnamefont
  {Oganesyan}}\ and\ \bibinfo {author} {\bibfnamefont {D.~A.}\ \bibnamefont
  {Huse}},\ }\bibfield  {title} {\enquote {\bibinfo {title} {Localization of
  interacting fermions at high temperature},}\ }\href {\doibase
  10.1103/PhysRevB.75.155111} {\bibfield  {journal} {\bibinfo  {journal} {Phys.
  Rev. B}\ }\textbf {\bibinfo {volume} {75}},\ \bibinfo {pages} {155111}
  (\bibinfo {year} {2007})}\BibitemShut {NoStop}%
\bibitem [{\citenamefont {Monthus}\ and\ \citenamefont
  {Garel}(2010)}]{monthus10}%
  \BibitemOpen
  \bibfield  {author} {\bibinfo {author} {\bibfnamefont {C.}~\bibnamefont
  {Monthus}}\ and\ \bibinfo {author} {\bibfnamefont {T.}~\bibnamefont
  {Garel}},\ }\bibfield  {title} {\enquote {\bibinfo {title} {Many-body
  localization transition in a lattice model of interacting fermions:
  Statistics of renormalized hoppings in configuration space},}\ }\href
  {\doibase 10.1103/PhysRevB.81.134202} {\bibfield  {journal} {\bibinfo
  {journal} {Phys. Rev. B}\ }\textbf {\bibinfo {volume} {81}},\ \bibinfo
  {pages} {134202} (\bibinfo {year} {2010})}\BibitemShut {NoStop}%
\bibitem [{\citenamefont {Luitz}\ \emph {et~al.}(2015)\citenamefont {Luitz},
  \citenamefont {Laflorencie},\ and\ \citenamefont {Alet}}]{luitz15}%
  \BibitemOpen
  \bibfield  {author} {\bibinfo {author} {\bibfnamefont {D.~J.}\ \bibnamefont
  {Luitz}}, \bibinfo {author} {\bibfnamefont {N.}~\bibnamefont {Laflorencie}},
  \ and\ \bibinfo {author} {\bibfnamefont {F.}~\bibnamefont {Alet}},\
  }\bibfield  {title} {\enquote {\bibinfo {title} {Many-body localization edge
  in the random-field {Heisenberg} chain},}\ }\href {\doibase
  10.1103/PhysRevB.91.081103} {\bibfield  {journal} {\bibinfo  {journal} {Phys.
  Rev. B}\ }\textbf {\bibinfo {volume} {91}},\ \bibinfo {pages} {081103}
  (\bibinfo {year} {2015})}\BibitemShut {NoStop}%
\bibitem [{\citenamefont {Andraschko}\ \emph {et~al.}(2014)\citenamefont
  {Andraschko}, \citenamefont {Enss},\ and\ \citenamefont {Sirker}}]{ZZZ5_4}%
  \BibitemOpen
  \bibfield  {author} {\bibinfo {author} {\bibfnamefont {F.}~\bibnamefont
  {Andraschko}}, \bibinfo {author} {\bibfnamefont {T.}~\bibnamefont {Enss}}, \
  and\ \bibinfo {author} {\bibfnamefont {J.}~\bibnamefont {Sirker}},\
  }\bibfield  {title} {\enquote {\bibinfo {title} {Purification and many-body
  localization in cold atomic gases},}\ }\href {\doibase
  10.1103/PhysRevLett.113.217201} {\bibfield  {journal} {\bibinfo  {journal}
  {Phys. Rev. Lett.}\ }\textbf {\bibinfo {volume} {113}},\ \bibinfo {pages}
  {217201} (\bibinfo {year} {2014})}\BibitemShut {NoStop}%
\bibitem [{\citenamefont {Ponte}\ \emph {et~al.}(2015)\citenamefont {Ponte},
  \citenamefont {Papi\ifmmode~\acute{c}\else \'{c}\fi{}}, \citenamefont
  {Huveneers},\ and\ \citenamefont {Abanin}}]{Ponte2015}%
  \BibitemOpen
  \bibfield  {author} {\bibinfo {author} {\bibfnamefont {P.}~\bibnamefont
  {Ponte}}, \bibinfo {author} {\bibfnamefont {Z.}~\bibnamefont
  {Papi\ifmmode~\acute{c}\else \'{c}\fi{}}}, \bibinfo {author} {\bibfnamefont
  {F.}~\bibnamefont {Huveneers}}, \ and\ \bibinfo {author} {\bibfnamefont
  {D.~A.}\ \bibnamefont {Abanin}},\ }\bibfield  {title} {\enquote {\bibinfo
  {title} {Many-body localization in periodically driven systems},}\ }\href
  {\doibase 10.1103/PhysRevLett.114.140401} {\bibfield  {journal} {\bibinfo
  {journal} {Phys. Rev. Lett.}\ }\textbf {\bibinfo {volume} {114}},\ \bibinfo
  {pages} {140401} (\bibinfo {year} {2015})}\BibitemShut {NoStop}%
\bibitem [{\citenamefont {Lazarides}\ \emph {et~al.}(2015)\citenamefont
  {Lazarides}, \citenamefont {Das},\ and\ \citenamefont
  {Moessner}}]{lazarides15}%
  \BibitemOpen
  \bibfield  {author} {\bibinfo {author} {\bibfnamefont {A.}~\bibnamefont
  {Lazarides}}, \bibinfo {author} {\bibfnamefont {A.}~\bibnamefont {Das}}, \
  and\ \bibinfo {author} {\bibfnamefont {R.}~\bibnamefont {Moessner}},\
  }\bibfield  {title} {\enquote {\bibinfo {title} {Fate of many-body
  localization under periodic driving},}\ }\href {\doibase
  10.1103/PhysRevLett.115.030402} {\bibfield  {journal} {\bibinfo  {journal}
  {Phys. Rev. Lett.}\ }\textbf {\bibinfo {volume} {115}},\ \bibinfo {pages}
  {030402} (\bibinfo {year} {2015})}\BibitemShut {NoStop}%
\bibitem [{\citenamefont {Vasseur}\ \emph {et~al.}(2015)\citenamefont
  {Vasseur}, \citenamefont {Parameswaran},\ and\ \citenamefont
  {Moore}}]{vasseur15a}%
  \BibitemOpen
  \bibfield  {author} {\bibinfo {author} {\bibfnamefont {R.}~\bibnamefont
  {Vasseur}}, \bibinfo {author} {\bibfnamefont {S.~A.}\ \bibnamefont
  {Parameswaran}}, \ and\ \bibinfo {author} {\bibfnamefont {J.~E.}\
  \bibnamefont {Moore}},\ }\bibfield  {title} {\enquote {\bibinfo {title}
  {Quantum revivals and many-body localization},}\ }\href {\doibase
  10.1103/PhysRevB.91.140202} {\bibfield  {journal} {\bibinfo  {journal} {Phys.
  Rev. B}\ }\textbf {\bibinfo {volume} {91}},\ \bibinfo {pages} {140202}
  (\bibinfo {year} {2015})}\BibitemShut {NoStop}%
\bibitem [{\citenamefont {Serbyn}\ \emph {et~al.}(2014)\citenamefont {Serbyn},
  \citenamefont {Papi\'{c}},\ and\ \citenamefont {Abanin}}]{serbyn2014a}%
  \BibitemOpen
  \bibfield  {author} {\bibinfo {author} {\bibfnamefont {M.}~\bibnamefont
  {Serbyn}}, \bibinfo {author} {\bibfnamefont {Z.}~\bibnamefont {Papi\'{c}}}, \
  and\ \bibinfo {author} {\bibfnamefont {D.~A.}\ \bibnamefont {Abanin}},\
  }\bibfield  {title} {\enquote {\bibinfo {title} {Quantum quenches in the
  many-body localized phase},}\ }\href {\doibase 10.1103/PhysRevB.90.174302}
  {\bibfield  {journal} {\bibinfo  {journal} {Phys. Rev. B}\ }\textbf {\bibinfo
  {volume} {90}},\ \bibinfo {pages} {174302} (\bibinfo {year}
  {2014})}\BibitemShut {NoStop}%
\bibitem [{\citenamefont {Pekker}\ \emph {et~al.}(2014)\citenamefont {Pekker},
  \citenamefont {Refael}, \citenamefont {Altman}, \citenamefont {Demler},\ and\
  \citenamefont {Oganesyan}}]{pekker2014}%
  \BibitemOpen
  \bibfield  {author} {\bibinfo {author} {\bibfnamefont {D.}~\bibnamefont
  {Pekker}}, \bibinfo {author} {\bibfnamefont {G.}~\bibnamefont {Refael}},
  \bibinfo {author} {\bibfnamefont {E.}~\bibnamefont {Altman}}, \bibinfo
  {author} {\bibfnamefont {E.}~\bibnamefont {Demler}}, \ and\ \bibinfo {author}
  {\bibfnamefont {V.}~\bibnamefont {Oganesyan}},\ }\bibfield  {title} {\enquote
  {\bibinfo {title} {Hilbert-glass transition: New universality of
  temperature-tuned many-body dynamical quantum criticality},}\ }\href
  {\doibase 10.1103/PhysRevX.4.011052} {\bibfield  {journal} {\bibinfo
  {journal} {Phys. Rev. X}\ }\textbf {\bibinfo {volume} {4}},\ \bibinfo {pages}
  {011052} (\bibinfo {year} {2014})}\BibitemShut {NoStop}%
\bibitem [{\citenamefont {Torres-Herrera}\ and\ \citenamefont
  {Santos}(2015)}]{torres15}%
  \BibitemOpen
  \bibfield  {author} {\bibinfo {author} {\bibfnamefont {E.~J.}\ \bibnamefont
  {Torres-Herrera}}\ and\ \bibinfo {author} {\bibfnamefont {Lea~F.}\
  \bibnamefont {Santos}},\ }\bibfield  {title} {\enquote {\bibinfo {title}
  {Dynamics at the many-body localization transition},}\ }\href {\doibase
  10.1103/PhysRevB.92.014208} {\bibfield  {journal} {\bibinfo  {journal} {Phys.
  Rev. B}\ }\textbf {\bibinfo {volume} {92}},\ \bibinfo {pages} {014208}
  (\bibinfo {year} {2015})}\BibitemShut {NoStop}%
\bibitem [{\citenamefont {T\'avora}\ \emph {et~al.}(2016)\citenamefont
  {T\'avora}, \citenamefont {Torres-Herrera},\ and\ \citenamefont
  {Santos}}]{torres16}%
  \BibitemOpen
  \bibfield  {author} {\bibinfo {author} {\bibfnamefont {M.}~\bibnamefont
  {T\'avora}}, \bibinfo {author} {\bibfnamefont {E.~J.}\ \bibnamefont
  {Torres-Herrera}}, \ and\ \bibinfo {author} {\bibfnamefont {L.~F.}\
  \bibnamefont {Santos}},\ }\bibfield  {title} {\enquote {\bibinfo {title}
  {Inevitable power-law behavior of isolated many-body quantum systems and how
  it anticipates thermalization},}\ }\href {\doibase
  10.1103/PhysRevA.94.041603} {\bibfield  {journal} {\bibinfo  {journal} {Phys.
  Rev. A}\ }\textbf {\bibinfo {volume} {94}},\ \bibinfo {pages} {041603}
  (\bibinfo {year} {2016})}\BibitemShut {NoStop}%
\bibitem [{\citenamefont {Laumann}\ \emph {et~al.}(2014)\citenamefont
  {Laumann}, \citenamefont {Pal},\ and\ \citenamefont
  {Scardicchio}}]{laumann2015}%
  \BibitemOpen
  \bibfield  {author} {\bibinfo {author} {\bibfnamefont {C.~R.}\ \bibnamefont
  {Laumann}}, \bibinfo {author} {\bibfnamefont {A.}~\bibnamefont {Pal}}, \ and\
  \bibinfo {author} {\bibfnamefont {A.}~\bibnamefont {Scardicchio}},\
  }\bibfield  {title} {\enquote {\bibinfo {title} {Many-body mobility edge in a
  mean-field quantum spin glass},}\ }\href {\doibase
  10.1103/PhysRevLett.113.200405} {\bibfield  {journal} {\bibinfo  {journal}
  {Phys. Rev. Lett.}\ }\textbf {\bibinfo {volume} {113}},\ \bibinfo {pages}
  {200405} (\bibinfo {year} {2014})}\BibitemShut {NoStop}%
\bibitem [{\citenamefont {Huse}\ \emph {et~al.}(2014)\citenamefont {Huse},
  \citenamefont {Nandkishore},\ and\ \citenamefont {Oganesyan}}]{huse14}%
  \BibitemOpen
  \bibfield  {author} {\bibinfo {author} {\bibfnamefont {D.~A.}\ \bibnamefont
  {Huse}}, \bibinfo {author} {\bibfnamefont {R.}~\bibnamefont {Nandkishore}}, \
  and\ \bibinfo {author} {\bibfnamefont {V.}~\bibnamefont {Oganesyan}},\
  }\bibfield  {title} {\enquote {\bibinfo {title} {Phenomenology of fully
  many-body-localized systems},}\ }\href {\doibase 10.1103/PhysRevB.90.174202}
  {\bibfield  {journal} {\bibinfo  {journal} {Phys. Rev. B}\ }\textbf {\bibinfo
  {volume} {90}},\ \bibinfo {pages} {174202} (\bibinfo {year}
  {2014})}\BibitemShut {NoStop}%
\bibitem [{\citenamefont {Gopalakrishnan}\ \emph {et~al.}(2017)\citenamefont
  {Gopalakrishnan}, \citenamefont {Islam},\ and\ \citenamefont
  {Knap}}]{gopal17}%
  \BibitemOpen
  \bibfield  {author} {\bibinfo {author} {\bibfnamefont {S.}~\bibnamefont
  {Gopalakrishnan}}, \bibinfo {author} {\bibfnamefont {K.~R.}\ \bibnamefont
  {Islam}}, \ and\ \bibinfo {author} {\bibfnamefont {M.}~\bibnamefont {Knap}},\
  }\bibfield  {title} {\enquote {\bibinfo {title} {Noise-induced subdiffusion
  in strongly localized quantum systems},}\ }\href {\doibase
  10.1103/PhysRevLett.119.046601} {\bibfield  {journal} {\bibinfo  {journal}
  {Phys. Rev. Lett.}\ }\textbf {\bibinfo {volume} {119}},\ \bibinfo {pages}
  {046601} (\bibinfo {year} {2017})}\BibitemShut {NoStop}%
\bibitem [{\citenamefont {Hauschild}\ \emph {et~al.}(2016)\citenamefont
  {Hauschild}, \citenamefont {Heidrich-Meisner},\ and\ \citenamefont
  {Pollmann}}]{Hauschild_2016}%
  \BibitemOpen
  \bibfield  {author} {\bibinfo {author} {\bibfnamefont {J.}~\bibnamefont
  {Hauschild}}, \bibinfo {author} {\bibfnamefont {F.}~\bibnamefont
  {Heidrich-Meisner}}, \ and\ \bibinfo {author} {\bibfnamefont
  {F.}~\bibnamefont {Pollmann}},\ }\bibfield  {title} {\enquote {\bibinfo
  {title} {Domain-wall melting as a probe of many-body localization},}\ }\href
  {\doibase 10.1103/physrevb.94.161109} {\bibfield  {journal} {\bibinfo
  {journal} {Physical Review B}\ }\textbf {\bibinfo {volume} {94}},\ \bibinfo
  {pages} {161109} (\bibinfo {year} {2016})}\BibitemShut {NoStop}%
\bibitem [{\citenamefont {Herbrych}\ \emph {et~al.}(2013)\citenamefont
  {Herbrych}, \citenamefont {Kokalj},\ and\ \citenamefont
  {Prelov\ifmmode~\check{s}\else \v{s}\fi{}ek}}]{herbrych13}%
  \BibitemOpen
  \bibfield  {author} {\bibinfo {author} {\bibfnamefont {J.}~\bibnamefont
  {Herbrych}}, \bibinfo {author} {\bibfnamefont {J.}~\bibnamefont {Kokalj}}, \
  and\ \bibinfo {author} {\bibfnamefont {P.}~\bibnamefont
  {Prelov\ifmmode~\check{s}\else \v{s}\fi{}ek}},\ }\bibfield  {title} {\enquote
  {\bibinfo {title} {Local spin relaxation within the random {Heisenberg}
  chain},}\ }\href {\doibase 10.1103/PhysRevLett.111.147203} {\bibfield
  {journal} {\bibinfo  {journal} {Phys. Rev. Lett.}\ }\textbf {\bibinfo
  {volume} {111}},\ \bibinfo {pages} {147203} (\bibinfo {year}
  {2013})}\BibitemShut {NoStop}%
\bibitem [{\citenamefont {Imbrie}(2016{\natexlab{a}})}]{imbrie16}%
  \BibitemOpen
  \bibfield  {author} {\bibinfo {author} {\bibfnamefont {J.~Z.}\ \bibnamefont
  {Imbrie}},\ }\bibfield  {title} {\enquote {\bibinfo {title} {Diagonalization
  and many-body localization for a disordered quantum spin chain},}\ }\href
  {\doibase 10.1103/PhysRevLett.117.027201} {\bibfield  {journal} {\bibinfo
  {journal} {Phys. Rev. Lett.}\ }\textbf {\bibinfo {volume} {117}},\ \bibinfo
  {pages} {027201} (\bibinfo {year} {2016}{\natexlab{a}})}\BibitemShut
  {NoStop}%
\bibitem [{\citenamefont {Steinigeweg}\ \emph {et~al.}(2016)\citenamefont
  {Steinigeweg}, \citenamefont {Herbrych}, \citenamefont {Pollmann},\ and\
  \citenamefont {Brenig}}]{steinigeweg16}%
  \BibitemOpen
  \bibfield  {author} {\bibinfo {author} {\bibfnamefont {R.}~\bibnamefont
  {Steinigeweg}}, \bibinfo {author} {\bibfnamefont {J.}~\bibnamefont
  {Herbrych}}, \bibinfo {author} {\bibfnamefont {F.}~\bibnamefont {Pollmann}},
  \ and\ \bibinfo {author} {\bibfnamefont {W.}~\bibnamefont {Brenig}},\
  }\bibfield  {title} {\enquote {\bibinfo {title} {Typicality approach to the
  optical conductivity in thermal and many-body localized phases},}\ }\href
  {\doibase 10.1103/PhysRevB.94.180401} {\bibfield  {journal} {\bibinfo
  {journal} {Phys. Rev. B}\ }\textbf {\bibinfo {volume} {94}},\ \bibinfo
  {pages} {180401} (\bibinfo {year} {2016})}\BibitemShut {NoStop}%
\bibitem [{\citenamefont {Herbrych}\ and\ \citenamefont
  {Kokalj}(2017)}]{Herbrych17}%
  \BibitemOpen
  \bibfield  {author} {\bibinfo {author} {\bibfnamefont {J.}~\bibnamefont
  {Herbrych}}\ and\ \bibinfo {author} {\bibfnamefont {J.}~\bibnamefont
  {Kokalj}},\ }\bibfield  {title} {\enquote {\bibinfo {title} {Effective
  realization of random magnetic fields in compounds with large single-ion
  anisotropy},}\ }\href {\doibase 10.1103/PhysRevB.95.125129} {\bibfield
  {journal} {\bibinfo  {journal} {Phys. Rev. B}\ }\textbf {\bibinfo {volume}
  {95}},\ \bibinfo {pages} {125129} (\bibinfo {year} {2017})}\BibitemShut
  {NoStop}%
\bibitem [{\citenamefont {Kondov}\ \emph {et~al.}(2015)\citenamefont {Kondov},
  \citenamefont {McGehee}, \citenamefont {Xu},\ and\ \citenamefont
  {DeMarco}}]{kondov15}%
  \BibitemOpen
  \bibfield  {author} {\bibinfo {author} {\bibfnamefont {S.~S.}\ \bibnamefont
  {Kondov}}, \bibinfo {author} {\bibfnamefont {W.~R.}\ \bibnamefont {McGehee}},
  \bibinfo {author} {\bibfnamefont {W.}~\bibnamefont {Xu}}, \ and\ \bibinfo
  {author} {\bibfnamefont {B.}~\bibnamefont {DeMarco}},\ }\bibfield  {title}
  {\enquote {\bibinfo {title} {Disorder-induced localization in a strongly
  correlated atomic {Hubbard} gas},}\ }\href {\doibase
  10.1103/PhysRevLett.114.083002} {\bibfield  {journal} {\bibinfo  {journal}
  {Phys. Rev. Lett.}\ }\textbf {\bibinfo {volume} {114}},\ \bibinfo {pages}
  {083002} (\bibinfo {year} {2015})}\BibitemShut {NoStop}%
\bibitem [{\citenamefont {Schreiber}\ \emph {et~al.}(2015)\citenamefont
  {Schreiber}, \citenamefont {Hodgman}, \citenamefont {Bordia}, \citenamefont
  {L{\"{u}}schen}, \citenamefont {Fischer}, \citenamefont {Vosk}, \citenamefont
  {Altman}, \citenamefont {Schneider},\ and\ \citenamefont
  {Bloch}}]{schreiber15}%
  \BibitemOpen
  \bibfield  {author} {\bibinfo {author} {\bibfnamefont {M.}~\bibnamefont
  {Schreiber}}, \bibinfo {author} {\bibfnamefont {S.~S.}\ \bibnamefont
  {Hodgman}}, \bibinfo {author} {\bibfnamefont {P.}~\bibnamefont {Bordia}},
  \bibinfo {author} {\bibfnamefont {H.~P.}\ \bibnamefont {L{\"{u}}schen}},
  \bibinfo {author} {\bibfnamefont {Mark~H}\ \bibnamefont {Fischer}}, \bibinfo
  {author} {\bibfnamefont {Ronen}\ \bibnamefont {Vosk}}, \bibinfo {author}
  {\bibfnamefont {Ehud}\ \bibnamefont {Altman}}, \bibinfo {author}
  {\bibfnamefont {Ulrich}\ \bibnamefont {Schneider}}, \ and\ \bibinfo {author}
  {\bibfnamefont {Immanuel}\ \bibnamefont {Bloch}},\ }\bibfield  {title}
  {\enquote {\bibinfo {title} {{Observation of many-body localization of
  interacting fermions in a quasi-random optical lattice}},}\ }\href {\doibase
  10.1126/science.aaa7432} {\bibfield  {journal} {\bibinfo  {journal}
  {Science}\ }\textbf {\bibinfo {volume} {349}},\ \bibinfo {pages} {842}
  (\bibinfo {year} {2015})}\BibitemShut {NoStop}%
\bibitem [{\citenamefont {Choi}\ \emph {et~al.}(2016)\citenamefont {Choi},
  \citenamefont {Hild}, \citenamefont {Zeiher}, \citenamefont {Schau{\ss}},
  \citenamefont {Rubio-Abadal}, \citenamefont {Yefsah}, \citenamefont
  {Khemani}, \citenamefont {Huse}, \citenamefont {Bloch},\ and\ \citenamefont
  {Gross}}]{choi16}%
  \BibitemOpen
  \bibfield  {author} {\bibinfo {author} {\bibfnamefont {J.-Y.}\ \bibnamefont
  {Choi}}, \bibinfo {author} {\bibfnamefont {S.}~\bibnamefont {Hild}}, \bibinfo
  {author} {\bibfnamefont {J.}~\bibnamefont {Zeiher}}, \bibinfo {author}
  {\bibfnamefont {P.}~\bibnamefont {Schau{\ss}}}, \bibinfo {author}
  {\bibfnamefont {A.}~\bibnamefont {Rubio-Abadal}}, \bibinfo {author}
  {\bibfnamefont {T.}~\bibnamefont {Yefsah}}, \bibinfo {author} {\bibfnamefont
  {V.}~\bibnamefont {Khemani}}, \bibinfo {author} {\bibfnamefont {D.~A.}\
  \bibnamefont {Huse}}, \bibinfo {author} {\bibfnamefont {I.}~\bibnamefont
  {Bloch}}, \ and\ \bibinfo {author} {\bibfnamefont {C.}~\bibnamefont
  {Gross}},\ }\bibfield  {title} {\enquote {\bibinfo {title} {{Exploring the
  many-body localization transition in two dimensions}},}\ }\href {\doibase
  10.1126/science.aaf8834} {\bibfield  {journal} {\bibinfo  {journal}
  {Science}\ }\textbf {\bibinfo {volume} {352}},\ \bibinfo {pages} {1547}
  (\bibinfo {year} {2016})}\BibitemShut {NoStop}%
\bibitem [{\citenamefont {Bordia}\ \emph {et~al.}(2016)\citenamefont {Bordia},
  \citenamefont {L{\"{u}}schen}, \citenamefont {Hodgman}, \citenamefont
  {Schreiber}, \citenamefont {Bloch},\ and\ \citenamefont
  {Schneider}}]{bordia16}%
  \BibitemOpen
  \bibfield  {author} {\bibinfo {author} {\bibfnamefont {P.}~\bibnamefont
  {Bordia}}, \bibinfo {author} {\bibfnamefont {H.~P.}\ \bibnamefont
  {L{\"{u}}schen}}, \bibinfo {author} {\bibfnamefont {S.~S.}\ \bibnamefont
  {Hodgman}}, \bibinfo {author} {\bibfnamefont {M.}~\bibnamefont {Schreiber}},
  \bibinfo {author} {\bibfnamefont {I.}~\bibnamefont {Bloch}}, \ and\ \bibinfo
  {author} {\bibfnamefont {U.}~\bibnamefont {Schneider}},\ }\bibfield  {title}
  {\enquote {\bibinfo {title} {{Coupling Identical 1D Many-Body Localized
  Systems}},}\ }\href {\doibase 10.1103/PhysRevLett.116.140401} {\bibfield
  {journal} {\bibinfo  {journal} {Phys. Rev. Lett.}\ }\textbf {\bibinfo
  {volume} {116}},\ \bibinfo {pages} {140401} (\bibinfo {year}
  {2016})}\BibitemShut {NoStop}%
\bibitem [{\citenamefont {Bordia}\ \emph {et~al.}(2017)\citenamefont {Bordia},
  \citenamefont {L\"uschen}, \citenamefont {Scherg}, \citenamefont
  {Gopalakrishnan}, \citenamefont {Knap}, \citenamefont {Schneider},\ and\
  \citenamefont {Bloch}}]{bordia2017_1}%
  \BibitemOpen
  \bibfield  {author} {\bibinfo {author} {\bibfnamefont {P.}~\bibnamefont
  {Bordia}}, \bibinfo {author} {\bibfnamefont {H.}~\bibnamefont {L\"uschen}},
  \bibinfo {author} {\bibfnamefont {S.}~\bibnamefont {Scherg}}, \bibinfo
  {author} {\bibfnamefont {S.}~\bibnamefont {Gopalakrishnan}}, \bibinfo
  {author} {\bibfnamefont {M.}~\bibnamefont {Knap}}, \bibinfo {author}
  {\bibfnamefont {U.}~\bibnamefont {Schneider}}, \ and\ \bibinfo {author}
  {\bibfnamefont {I.}~\bibnamefont {Bloch}},\ }\bibfield  {title} {\enquote
  {\bibinfo {title} {Probing slow relaxation and many-body localization in
  two-dimensional quasiperiodic systems},}\ }\href {\doibase
  10.1103/PhysRevX.7.041047} {\bibfield  {journal} {\bibinfo  {journal} {Phys.
  Rev. X}\ }\textbf {\bibinfo {volume} {7}},\ \bibinfo {pages} {041047}
  (\bibinfo {year} {2017})}\BibitemShut {NoStop}%
\bibitem [{\citenamefont {Smith}\ \emph {et~al.}(2016)\citenamefont {Smith},
  \citenamefont {Lee}, \citenamefont {Richerme}, \citenamefont {Neyenhuis},
  \citenamefont {Hess}, \citenamefont {Hauke}, \citenamefont {Heyl},
  \citenamefont {Huse},\ and\ \citenamefont {Monroe}}]{smith2016}%
  \BibitemOpen
  \bibfield  {author} {\bibinfo {author} {\bibfnamefont {J.}~\bibnamefont
  {Smith}}, \bibinfo {author} {\bibfnamefont {A.}~\bibnamefont {Lee}}, \bibinfo
  {author} {\bibfnamefont {P.}~\bibnamefont {Richerme}}, \bibinfo {author}
  {\bibfnamefont {B.}~\bibnamefont {Neyenhuis}}, \bibinfo {author}
  {\bibfnamefont {P.~W.}\ \bibnamefont {Hess}}, \bibinfo {author}
  {\bibfnamefont {P.}~\bibnamefont {Hauke}}, \bibinfo {author} {\bibfnamefont
  {M.}~\bibnamefont {Heyl}}, \bibinfo {author} {\bibfnamefont {D.~A.}\
  \bibnamefont {Huse}}, \ and\ \bibinfo {author} {\bibfnamefont
  {C.}~\bibnamefont {Monroe}},\ }\bibfield  {title} {\enquote {\bibinfo {title}
  {Many-body localization in a quantum simulator with programmable random
  disorder},}\ }\href {\doibase dx.doi.org/10.1038/nphys3783} {\bibfield
  {journal} {\bibinfo  {journal} {Nat. Phys.}\ }\textbf {\bibinfo {volume}
  {12}},\ \bibinfo {pages} {907} (\bibinfo {year} {2016})}\BibitemShut
  {NoStop}%
\bibitem [{\citenamefont {\v{Z}nidari\v{c}}\ \emph {et~al.}(2008)\citenamefont
  {\v{Z}nidari\v{c}}, \citenamefont {Prosen},\ and\ \citenamefont
  {Prelov\ifmmode~\check{s}\else \v{s}\fi{}ek}}]{znidaric08}%
  \BibitemOpen
  \bibfield  {author} {\bibinfo {author} {\bibfnamefont {M.}~\bibnamefont
  {\v{Z}nidari\v{c}}}, \bibinfo {author} {\bibfnamefont {T.}~\bibnamefont
  {Prosen}}, \ and\ \bibinfo {author} {\bibfnamefont {P.}~\bibnamefont
  {Prelov\ifmmode~\check{s}\else \v{s}\fi{}ek}},\ }\bibfield  {title} {\enquote
  {\bibinfo {title} {Many-body localization in the {Heisenberg XXZ} magnet in a
  random field},}\ }\href {\doibase 10.1103/PhysRevB.77.064426} {\bibfield
  {journal} {\bibinfo  {journal} {Phys. Rev. B}\ }\textbf {\bibinfo {volume}
  {77}},\ \bibinfo {pages} {064426} (\bibinfo {year} {2008})}\BibitemShut
  {NoStop}%
\bibitem [{\citenamefont {Bardarson}\ \emph {et~al.}(2012)\citenamefont
  {Bardarson}, \citenamefont {Pollmann},\ and\ \citenamefont
  {Moore}}]{bardarson12}%
  \BibitemOpen
  \bibfield  {author} {\bibinfo {author} {\bibfnamefont {J.~H}\ \bibnamefont
  {Bardarson}}, \bibinfo {author} {\bibfnamefont {F.}~\bibnamefont {Pollmann}},
  \ and\ \bibinfo {author} {\bibfnamefont {J.~E.}\ \bibnamefont {Moore}},\
  }\bibfield  {title} {\enquote {\bibinfo {title} {Unbounded growth of
  entanglement in models of many-body localization},}\ }\href {\doibase
  10.1103/PhysRevLett.109.017202} {\bibfield  {journal} {\bibinfo  {journal}
  {Phys. Rev. Lett.}\ }\textbf {\bibinfo {volume} {109}},\ \bibinfo {pages}
  {017202} (\bibinfo {year} {2012})}\BibitemShut {NoStop}%
\bibitem [{\citenamefont {Kj{\"{a}}ll}\ \emph {et~al.}(2014)\citenamefont
  {Kj{\"{a}}ll}, \citenamefont {Bardarson},\ and\ \citenamefont
  {Pollmann}}]{kjall14}%
  \BibitemOpen
  \bibfield  {author} {\bibinfo {author} {\bibfnamefont {J.~A.}\ \bibnamefont
  {Kj{\"{a}}ll}}, \bibinfo {author} {\bibfnamefont {J.~H.}\ \bibnamefont
  {Bardarson}}, \ and\ \bibinfo {author} {\bibfnamefont {F.}~\bibnamefont
  {Pollmann}},\ }\bibfield  {title} {\enquote {\bibinfo {title} {Many-body
  localization in a disordered quantum {Ising} chain},}\ }\href {\doibase
  10.1103/PhysRevLett.113.107204} {\ \textbf {\bibinfo {volume} {113}},\
  \bibinfo {pages} {107204} (\bibinfo {year} {2014})}\BibitemShut {NoStop}%
\bibitem [{\citenamefont {Serbyn}\ \emph {et~al.}(2015)\citenamefont {Serbyn},
  \citenamefont {Papi{\'{c}}},\ and\ \citenamefont {Abanin}}]{serbyn15}%
  \BibitemOpen
  \bibfield  {author} {\bibinfo {author} {\bibfnamefont {M.}~\bibnamefont
  {Serbyn}}, \bibinfo {author} {\bibfnamefont {Z.}~\bibnamefont {Papi{\'{c}}}},
  \ and\ \bibinfo {author} {\bibfnamefont {D.~A.}\ \bibnamefont {Abanin}},\
  }\bibfield  {title} {\enquote {\bibinfo {title} {Criterion for many-body
  localization-delocalization phase transition},}\ }\href {\doibase
  10.1103/PhysRevX.5.041047} {\bibfield  {journal} {\bibinfo  {journal} {Phys.
  Rev. X}\ }\textbf {\bibinfo {volume} {5}},\ \bibinfo {pages} {041047}
  (\bibinfo {year} {2015})}\BibitemShut {NoStop}%
\bibitem [{\citenamefont {Luitz}\ \emph {et~al.}(2016)\citenamefont {Luitz},
  \citenamefont {Laflorencie},\ and\ \citenamefont {Alet}}]{luitz16}%
  \BibitemOpen
  \bibfield  {author} {\bibinfo {author} {\bibfnamefont {D.~J.}\ \bibnamefont
  {Luitz}}, \bibinfo {author} {\bibfnamefont {N.}~\bibnamefont {Laflorencie}},
  \ and\ \bibinfo {author} {\bibfnamefont {F.}~\bibnamefont {Alet}},\
  }\bibfield  {title} {\enquote {\bibinfo {title} {{Extended slow dynamical
  regime prefiguring the many-body localization transition}},}\ }\href
  {\doibase 10.1103/PhysRevB.93.060201} {\bibfield  {journal} {\bibinfo
  {journal} {Phys. Rev. B}\ }\textbf {\bibinfo {volume} {93}},\ \bibinfo
  {pages} {060201(R)} (\bibinfo {year} {2016})}\BibitemShut {NoStop}%
\bibitem [{\citenamefont {Serbyn}\ \emph {et~al.}(2013)\citenamefont {Serbyn},
  \citenamefont {Papi\'{c}},\ and\ \citenamefont {Abanin}}]{serbyn13_1}%
  \BibitemOpen
  \bibfield  {author} {\bibinfo {author} {\bibfnamefont {M.}~\bibnamefont
  {Serbyn}}, \bibinfo {author} {\bibfnamefont {Z.}~\bibnamefont {Papi\'{c}}}, \
  and\ \bibinfo {author} {\bibfnamefont {D.~A.}\ \bibnamefont {Abanin}},\
  }\bibfield  {title} {\enquote {\bibinfo {title} {Universal slow growth of
  entanglement in interacting strongly disordered systems},}\ }\href {\doibase
  10.1103/PhysRevLett.110.260601} {\bibfield  {journal} {\bibinfo  {journal}
  {Phys. Rev. Lett.}\ }\textbf {\bibinfo {volume} {110}},\ \bibinfo {pages}
  {260601} (\bibinfo {year} {2013})}\BibitemShut {NoStop}%
\bibitem [{\citenamefont {Bera}\ \emph {et~al.}(2015)\citenamefont {Bera},
  \citenamefont {Schomerus}, \citenamefont {Heidrich-Meisner},\ and\
  \citenamefont {Bardarson}}]{bera15}%
  \BibitemOpen
  \bibfield  {author} {\bibinfo {author} {\bibfnamefont {S.}~\bibnamefont
  {Bera}}, \bibinfo {author} {\bibfnamefont {H.}~\bibnamefont {Schomerus}},
  \bibinfo {author} {\bibfnamefont {F.}~\bibnamefont {Heidrich-Meisner}}, \
  and\ \bibinfo {author} {\bibfnamefont {J.~H.}\ \bibnamefont {Bardarson}},\
  }\bibfield  {title} {\enquote {\bibinfo {title} {Many-body localization
  characterized from a one-particle perspective},}\ }\href {\doibase
  10.1103/PhysRevLett.115.046603} {\bibfield  {journal} {\bibinfo  {journal}
  {Phys. Rev. Lett.}\ }\textbf {\bibinfo {volume} {115}},\ \bibinfo {pages}
  {046603} (\bibinfo {year} {2015})}\BibitemShut {NoStop}%
\bibitem [{\citenamefont {Altman}\ and\ \citenamefont {Vosk}(2015)}]{altman15}%
  \BibitemOpen
  \bibfield  {author} {\bibinfo {author} {\bibfnamefont {E.}~\bibnamefont
  {Altman}}\ and\ \bibinfo {author} {\bibfnamefont {R.}~\bibnamefont {Vosk}},\
  }\bibfield  {title} {\enquote {\bibinfo {title} {Universal dynamics and
  renormalization in many-body-localized systems},}\ }\href {\doibase
  10.1146/annurev-conmatphys-031214-014701} {\bibfield  {journal} {\bibinfo
  {journal} {Annu. Rev. Condens. Matter Phys.}\ }\textbf {\bibinfo {volume}
  {6}},\ \bibinfo {pages} {383} (\bibinfo {year} {2015})}\BibitemShut {NoStop}%
\bibitem [{\citenamefont {Agarwal}\ \emph {et~al.}(2015)\citenamefont
  {Agarwal}, \citenamefont {Gopalakrishnan}, \citenamefont {Knap},
  \citenamefont {M\"uller},\ and\ \citenamefont {Demler}}]{agarwal15}%
  \BibitemOpen
  \bibfield  {author} {\bibinfo {author} {\bibfnamefont {K.}~\bibnamefont
  {Agarwal}}, \bibinfo {author} {\bibfnamefont {S.}~\bibnamefont
  {Gopalakrishnan}}, \bibinfo {author} {\bibfnamefont {M.}~\bibnamefont
  {Knap}}, \bibinfo {author} {\bibfnamefont {M.}~\bibnamefont {M\"uller}}, \
  and\ \bibinfo {author} {\bibfnamefont {E.}~\bibnamefont {Demler}},\
  }\bibfield  {title} {\enquote {\bibinfo {title} {Anomalous diffusion and
  {Griffiths} effects near the many-body localization transition},}\ }\href
  {\doibase 10.1103/PhysRevLett.114.160401} {\bibfield  {journal} {\bibinfo
  {journal} {Phys. Rev. Lett.}\ }\textbf {\bibinfo {volume} {114}},\ \bibinfo
  {pages} {160401} (\bibinfo {year} {2015})}\BibitemShut {NoStop}%
\bibitem [{\citenamefont {Gopalakrishnan}\ \emph {et~al.}(2015)\citenamefont
  {Gopalakrishnan}, \citenamefont {M\"uller}, \citenamefont {Khemani},
  \citenamefont {Knap}, \citenamefont {Demler},\ and\ \citenamefont
  {Huse}}]{gopal15}%
  \BibitemOpen
  \bibfield  {author} {\bibinfo {author} {\bibfnamefont {S.}~\bibnamefont
  {Gopalakrishnan}}, \bibinfo {author} {\bibfnamefont {M.}~\bibnamefont
  {M\"uller}}, \bibinfo {author} {\bibfnamefont {V.}~\bibnamefont {Khemani}},
  \bibinfo {author} {\bibfnamefont {M.}~\bibnamefont {Knap}}, \bibinfo {author}
  {\bibfnamefont {E.}~\bibnamefont {Demler}}, \ and\ \bibinfo {author}
  {\bibfnamefont {D.~A.}\ \bibnamefont {Huse}},\ }\bibfield  {title} {\enquote
  {\bibinfo {title} {Low-frequency conductivity in many-body localized
  systems},}\ }\href {\doibase 10.1103/PhysRevB.92.104202} {\bibfield
  {journal} {\bibinfo  {journal} {Phys. Rev. B}\ }\textbf {\bibinfo {volume}
  {92}},\ \bibinfo {pages} {104202} (\bibinfo {year} {2015})}\BibitemShut
  {NoStop}%
\bibitem [{\citenamefont {{\v Znidari\v c}}\ \emph {et~al.}(2016)\citenamefont
  {{\v Znidari\v c}}, \citenamefont {Scardicchio},\ and\ \citenamefont
  {Varma}}]{znidaric16}%
  \BibitemOpen
  \bibfield  {author} {\bibinfo {author} {\bibfnamefont {M.}~\bibnamefont {{\v
  Znidari\v c}}}, \bibinfo {author} {\bibfnamefont {A.}~\bibnamefont
  {Scardicchio}}, \ and\ \bibinfo {author} {\bibfnamefont {V.~K.}\ \bibnamefont
  {Varma}},\ }\bibfield  {title} {\enquote {\bibinfo {title} {Diffusive and
  subdiffusive spin transport in the ergodic phase of a many-body localizable
  system},}\ }\href {http://dx.doi.org/10.1103/PhysRevLett.117.040601}
  {\bibfield  {journal} {\bibinfo  {journal} {Phys. Rev. Lett.}\ }\textbf
  {\bibinfo {volume} {117}},\ \bibinfo {pages} {040601} (\bibinfo {year}
  {2016})}\BibitemShut {NoStop}%
\bibitem [{\citenamefont {Mierzejewski}\ \emph {et~al.}(2016)\citenamefont
  {Mierzejewski}, \citenamefont {Herbrych},\ and\ \citenamefont
  {Prelov\ifmmode~\check{s}\else \v{s}\fi{}ek}}]{mierzejewski2016}%
  \BibitemOpen
  \bibfield  {author} {\bibinfo {author} {\bibfnamefont {M.}~\bibnamefont
  {Mierzejewski}}, \bibinfo {author} {\bibfnamefont {J.}~\bibnamefont
  {Herbrych}}, \ and\ \bibinfo {author} {\bibfnamefont {P.}~\bibnamefont
  {Prelov\ifmmode~\check{s}\else \v{s}\fi{}ek}},\ }\bibfield  {title} {\enquote
  {\bibinfo {title} {Universal dynamics of density correlations at the
  transition to the many-body localized state},}\ }\href {\doibase
  10.1103/PhysRevB.94.224207} {\bibfield  {journal} {\bibinfo  {journal} {Phys.
  Rev. B}\ }\textbf {\bibinfo {volume} {94}},\ \bibinfo {pages} {224207}
  (\bibinfo {year} {2016})}\BibitemShut {NoStop}%
\bibitem [{\citenamefont {Bar~Lev}\ and\ \citenamefont
  {Reichman}(2014)}]{lev14}%
  \BibitemOpen
  \bibfield  {author} {\bibinfo {author} {\bibfnamefont {Y.}~\bibnamefont
  {Bar~Lev}}\ and\ \bibinfo {author} {\bibfnamefont {D.~R.}\ \bibnamefont
  {Reichman}},\ }\bibfield  {title} {\enquote {\bibinfo {title} {Dynamics of
  many-body localization},}\ }\href {\doibase 10.1103/PhysRevB.89.220201}
  {\bibfield  {journal} {\bibinfo  {journal} {Phys. Rev. B}\ }\textbf {\bibinfo
  {volume} {89}},\ \bibinfo {pages} {220201} (\bibinfo {year}
  {2014})}\BibitemShut {NoStop}%
\bibitem [{\citenamefont {Bar~Lev}\ \emph {et~al.}(2015)\citenamefont
  {Bar~Lev}, \citenamefont {Cohen},\ and\ \citenamefont {Reichman}}]{lev15}%
  \BibitemOpen
  \bibfield  {author} {\bibinfo {author} {\bibfnamefont {Y.}~\bibnamefont
  {Bar~Lev}}, \bibinfo {author} {\bibfnamefont {G.}~\bibnamefont {Cohen}}, \
  and\ \bibinfo {author} {\bibfnamefont {D.~R.}\ \bibnamefont {Reichman}},\
  }\bibfield  {title} {\enquote {\bibinfo {title} {Absence of diffusion in an
  interacting system of spinless fermions on a one-dimensional disordered
  lattice},}\ }\href {\doibase 10.1103/PhysRevLett.114.100601} {\bibfield
  {journal} {\bibinfo  {journal} {Phys. Rev. Lett.}\ }\textbf {\bibinfo
  {volume} {114}},\ \bibinfo {pages} {100601} (\bibinfo {year}
  {2015})}\BibitemShut {NoStop}%
\bibitem [{\citenamefont {Bari\ifmmode \check{s}\else
  \v{s}\fi{}i\ifmmode~\acute{c}\else \'{c}\fi{}}\ \emph
  {et~al.}(2016)\citenamefont {Bari\ifmmode \check{s}\else
  \v{s}\fi{}i\ifmmode~\acute{c}\else \'{c}\fi{}}, \citenamefont {Kokalj},
  \citenamefont {Balog},\ and\ \citenamefont {Prelov\ifmmode~\check{s}\else
  \v{s}\fi{}ek}}]{barisic16}%
  \BibitemOpen
  \bibfield  {author} {\bibinfo {author} {\bibfnamefont {O.~S.}\ \bibnamefont
  {Bari\ifmmode \check{s}\else \v{s}\fi{}i\ifmmode~\acute{c}\else \'{c}\fi{}}},
  \bibinfo {author} {\bibfnamefont {J.}~\bibnamefont {Kokalj}}, \bibinfo
  {author} {\bibfnamefont {I.}~\bibnamefont {Balog}}, \ and\ \bibinfo {author}
  {\bibfnamefont {P.}~\bibnamefont {Prelov\ifmmode~\check{s}\else
  \v{s}\fi{}ek}},\ }\bibfield  {title} {\enquote {\bibinfo {title} {Dynamical
  conductivity and its fluctuations along the crossover to many-body
  localization},}\ }\href {\doibase 10.1103/PhysRevB.94.045126} {\bibfield
  {journal} {\bibinfo  {journal} {Phys. Rev. B}\ }\textbf {\bibinfo {volume}
  {94}},\ \bibinfo {pages} {045126} (\bibinfo {year} {2016})}\BibitemShut
  {NoStop}%
\bibitem [{\citenamefont {Bon\ifmmode~\check{c}\else \v{c}\fi{}a}\ and\
  \citenamefont {Mierzejewski}(2017)}]{bonca17}%
  \BibitemOpen
  \bibfield  {author} {\bibinfo {author} {\bibfnamefont {J.}~\bibnamefont
  {Bon\ifmmode~\check{c}\else \v{c}\fi{}a}}\ and\ \bibinfo {author}
  {\bibfnamefont {M.}~\bibnamefont {Mierzejewski}},\ }\bibfield  {title}
  {\enquote {\bibinfo {title} {Delocalized carriers in the {t-J} model with
  strong charge disorder},}\ }\href {\doibase 10.1103/PhysRevB.95.214201}
  {\bibfield  {journal} {\bibinfo  {journal} {Phys. Rev. B}\ }\textbf {\bibinfo
  {volume} {95}},\ \bibinfo {pages} {214201} (\bibinfo {year}
  {2017})}\BibitemShut {NoStop}%
\bibitem [{\citenamefont {Sierant}\ \emph {et~al.}(2017)\citenamefont
  {Sierant}, \citenamefont {Delande},\ and\ \citenamefont
  {Zakrzewski}}]{zakrzewski16}%
  \BibitemOpen
  \bibfield  {author} {\bibinfo {author} {\bibfnamefont {P.}~\bibnamefont
  {Sierant}}, \bibinfo {author} {\bibfnamefont {D.}~\bibnamefont {Delande}}, \
  and\ \bibinfo {author} {\bibfnamefont {J.}~\bibnamefont {Zakrzewski}},\
  }\bibfield  {title} {\enquote {\bibinfo {title} {Many-body localization due
  to random interactions},}\ }\href {\doibase 10.1103/PhysRevA.95.021601}
  {\bibfield  {journal} {\bibinfo  {journal} {Phys. Rev. A}\ }\textbf {\bibinfo
  {volume} {95}},\ \bibinfo {pages} {021601} (\bibinfo {year}
  {2017})}\BibitemShut {NoStop}%
\bibitem [{\citenamefont {Protopopov}\ and\ \citenamefont
  {Abanin}(2019)}]{protopopov2018}%
  \BibitemOpen
  \bibfield  {author} {\bibinfo {author} {\bibfnamefont {Ivan~V.}\ \bibnamefont
  {Protopopov}}\ and\ \bibinfo {author} {\bibfnamefont {Dmitry~A.}\
  \bibnamefont {Abanin}},\ }\bibfield  {title} {\enquote {\bibinfo {title}
  {Spin-mediated particle transport in the disordered {Hubbard} model},}\
  }\href {\doibase 10.1103/PhysRevB.99.115111} {\bibfield  {journal} {\bibinfo
  {journal} {Phys. Rev. B}\ }\textbf {\bibinfo {volume} {99}},\ \bibinfo
  {pages} {115111} (\bibinfo {year} {2019})}\BibitemShut {NoStop}%
\bibitem [{\citenamefont {Schecter}\ \emph {et~al.}(2018)\citenamefont
  {Schecter}, \citenamefont {Iadecola},\ and\ \citenamefont
  {Das~Sarma}}]{sankar2018}%
  \BibitemOpen
  \bibfield  {author} {\bibinfo {author} {\bibfnamefont {M.}~\bibnamefont
  {Schecter}}, \bibinfo {author} {\bibfnamefont {T.}~\bibnamefont {Iadecola}},
  \ and\ \bibinfo {author} {\bibfnamefont {S.}~\bibnamefont {Das~Sarma}},\
  }\bibfield  {title} {\enquote {\bibinfo {title} {Configuration-controlled
  many-body localization and the mobility emulsion},}\ }\href {\doibase
  10.1103/PhysRevB.98.174201} {\bibfield  {journal} {\bibinfo  {journal} {Phys.
  Rev. B}\ }\textbf {\bibinfo {volume} {98}},\ \bibinfo {pages} {174201}
  (\bibinfo {year} {2018})}\BibitemShut {NoStop}%
\bibitem [{\citenamefont {Zakrzewski}\ and\ \citenamefont
  {Delande}(2018)}]{zakrzewski2018}%
  \BibitemOpen
  \bibfield  {author} {\bibinfo {author} {\bibfnamefont {J.}~\bibnamefont
  {Zakrzewski}}\ and\ \bibinfo {author} {\bibfnamefont {D.}~\bibnamefont
  {Delande}},\ }\bibfield  {title} {\enquote {\bibinfo {title} {Spin-charge
  separation and many-body localization},}\ }\href {\doibase
  10.1103/PhysRevB.98.014203} {\bibfield  {journal} {\bibinfo  {journal} {Phys.
  Rev. B}\ }\textbf {\bibinfo {volume} {98}},\ \bibinfo {pages} {014203}
  (\bibinfo {year} {2018})}\BibitemShut {NoStop}%
\bibitem [{\citenamefont {Chandran}\ \emph {et~al.}(2014)\citenamefont
  {Chandran}, \citenamefont {Khemani}, \citenamefont {Laumann},\ and\
  \citenamefont {Sondhi}}]{Chandran2014}%
  \BibitemOpen
  \bibfield  {author} {\bibinfo {author} {\bibfnamefont {A.}~\bibnamefont
  {Chandran}}, \bibinfo {author} {\bibfnamefont {V.}~\bibnamefont {Khemani}},
  \bibinfo {author} {\bibfnamefont {C.~R.}\ \bibnamefont {Laumann}}, \ and\
  \bibinfo {author} {\bibfnamefont {S.~L.}\ \bibnamefont {Sondhi}},\ }\bibfield
   {title} {\enquote {\bibinfo {title} {Many-body localization and
  symmetry-protected topological order},}\ }\href
  {http://journals.aps.org/prb/abstract/10.1103/PhysRevB.89.144201} {\bibfield
  {journal} {\bibinfo  {journal} {Phys. Rev. B}\ }\textbf {\bibinfo {volume}
  {89}},\ \bibinfo {pages} {144201} (\bibinfo {year} {2014})}\BibitemShut
  {NoStop}%
\bibitem [{\citenamefont {Potter}\ and\ \citenamefont
  {Vasseur}(2016)}]{potter16}%
  \BibitemOpen
  \bibfield  {author} {\bibinfo {author} {\bibfnamefont {Andrew~C.}\
  \bibnamefont {Potter}}\ and\ \bibinfo {author} {\bibfnamefont {Romain}\
  \bibnamefont {Vasseur}},\ }\bibfield  {title} {\enquote {\bibinfo {title}
  {Symmetry constraints on many-body localization},}\ }\href {\doibase
  10.1103/PhysRevB.94.224206} {\bibfield  {journal} {\bibinfo  {journal} {Phys.
  Rev. B}\ }\textbf {\bibinfo {volume} {94}},\ \bibinfo {pages} {224206}
  (\bibinfo {year} {2016})}\BibitemShut {NoStop}%
\bibitem [{\citenamefont {Prelov\ifmmode~\check{s}\else \v{s}\fi{}ek}\ \emph
  {et~al.}(2016{\natexlab{a}})\citenamefont {Prelov\ifmmode~\check{s}\else
  \v{s}\fi{}ek}, \citenamefont {Bari\ifmmode \check{s}\else
  \v{s}\fi{}i\ifmmode~\acute{c}\else \'{c}\fi{}},\ and\ \citenamefont {\ifmmode
  \check{Z}\else \v{Z}\fi{}nidari\ifmmode~\check{c}\else
  \v{c}\fi{}}}]{prelovsek16}%
  \BibitemOpen
  \bibfield  {author} {\bibinfo {author} {\bibfnamefont {P.}~\bibnamefont
  {Prelov\ifmmode~\check{s}\else \v{s}\fi{}ek}}, \bibinfo {author}
  {\bibfnamefont {O.~S.}\ \bibnamefont {Bari\ifmmode \check{s}\else
  \v{s}\fi{}i\ifmmode~\acute{c}\else \'{c}\fi{}}}, \ and\ \bibinfo {author}
  {\bibfnamefont {M.}~\bibnamefont {\ifmmode \check{Z}\else
  \v{Z}\fi{}nidari\ifmmode~\check{c}\else \v{c}\fi{}}},\ }\bibfield  {title}
  {\enquote {\bibinfo {title} {Absence of full many-body localization in the
  disordered {Hubbard} chain},}\ }\href {\doibase 10.1103/PhysRevB.94.241104}
  {\bibfield  {journal} {\bibinfo  {journal} {Phys. Rev. B}\ }\textbf {\bibinfo
  {volume} {94}},\ \bibinfo {pages} {241104} (\bibinfo {year}
  {2016}{\natexlab{a}})}\BibitemShut {NoStop}%
\bibitem [{\citenamefont {Protopopov}\ \emph {et~al.}(2017)\citenamefont
  {Protopopov}, \citenamefont {Ho},\ and\ \citenamefont {Abanin}}]{proto2017}%
  \BibitemOpen
  \bibfield  {author} {\bibinfo {author} {\bibfnamefont {Ivan~V.}\ \bibnamefont
  {Protopopov}}, \bibinfo {author} {\bibfnamefont {Wen~Wei}\ \bibnamefont
  {Ho}}, \ and\ \bibinfo {author} {\bibfnamefont {Dmitry~A.}\ \bibnamefont
  {Abanin}},\ }\bibfield  {title} {\enquote {\bibinfo {title} {Effect of su(2)
  symmetry on many-body localization and thermalization},}\ }\href {\doibase
  10.1103/PhysRevB.96.041122} {\bibfield  {journal} {\bibinfo  {journal} {Phys.
  Rev. B}\ }\textbf {\bibinfo {volume} {96}},\ \bibinfo {pages} {041122}
  (\bibinfo {year} {2017})}\BibitemShut {NoStop}%
\bibitem [{\citenamefont {Friedman}\ \emph {et~al.}(2018)\citenamefont
  {Friedman}, \citenamefont {Vasseur}, \citenamefont {Potter},\ and\
  \citenamefont {Parameswaran}}]{friedman2017}%
  \BibitemOpen
  \bibfield  {author} {\bibinfo {author} {\bibfnamefont {Aaron~J.}\
  \bibnamefont {Friedman}}, \bibinfo {author} {\bibfnamefont {Romain}\
  \bibnamefont {Vasseur}}, \bibinfo {author} {\bibfnamefont {Andrew~C.}\
  \bibnamefont {Potter}}, \ and\ \bibinfo {author} {\bibfnamefont {S.~A.}\
  \bibnamefont {Parameswaran}},\ }\bibfield  {title} {\enquote {\bibinfo
  {title} {Localization-protected order in spin chains with non-abelian
  discrete symmetries},}\ }\href {\doibase 10.1103/PhysRevB.98.064203}
  {\bibfield  {journal} {\bibinfo  {journal} {Phys. Rev. B}\ }\textbf {\bibinfo
  {volume} {98}},\ \bibinfo {pages} {064203} (\bibinfo {year}
  {2018})}\BibitemShut {NoStop}%
\bibitem [{\citenamefont {Luitz}\ and\ \citenamefont
  {Bar~Lev}(2016{\natexlab{a}})}]{luitz2016prl}%
  \BibitemOpen
  \bibfield  {author} {\bibinfo {author} {\bibfnamefont {D.~J.}\ \bibnamefont
  {Luitz}}\ and\ \bibinfo {author} {\bibfnamefont {Y.}~\bibnamefont
  {Bar~Lev}},\ }\bibfield  {title} {\enquote {\bibinfo {title} {Anomalous
  thermalization in ergodic systems},}\ }\href {\doibase
  10.1103/PhysRevLett.117.170404} {\bibfield  {journal} {\bibinfo  {journal}
  {Phys. Rev. Lett.}\ }\textbf {\bibinfo {volume} {117}},\ \bibinfo {pages}
  {170404} (\bibinfo {year} {2016}{\natexlab{a}})}\BibitemShut {NoStop}%
\bibitem [{\citenamefont {Luitz}\ and\ \citenamefont
  {Bar~Lev}(2016{\natexlab{b}})}]{luitz116}%
  \BibitemOpen
  \bibfield  {author} {\bibinfo {author} {\bibfnamefont {D.~J.}\ \bibnamefont
  {Luitz}}\ and\ \bibinfo {author} {\bibfnamefont {Y.}~\bibnamefont
  {Bar~Lev}},\ }\bibfield  {title} {\enquote {\bibinfo {title} {The ergodic
  side of the many‐body localization transition},}\ }\href {\doibase
  10.1002/andp.201600350} {\bibfield  {journal} {\bibinfo  {journal} {Annalen
  der Physik}\ }\textbf {\bibinfo {volume} {529}},\ \bibinfo {pages} {1600350}
  (\bibinfo {year} {2016}{\natexlab{b}})}\BibitemShut {NoStop}%
\bibitem [{\citenamefont {Kozarzewski}\ \emph {et~al.}(2018)\citenamefont
  {Kozarzewski}, \citenamefont {Prelov\ifmmode~\check{s}\else \v{s}\fi{}ek},\
  and\ \citenamefont {Mierzejewski}}]{kozarzewski18}%
  \BibitemOpen
  \bibfield  {author} {\bibinfo {author} {\bibfnamefont {M.}~\bibnamefont
  {Kozarzewski}}, \bibinfo {author} {\bibfnamefont {P.}~\bibnamefont
  {Prelov\ifmmode~\check{s}\else \v{s}\fi{}ek}}, \ and\ \bibinfo {author}
  {\bibfnamefont {M.}~\bibnamefont {Mierzejewski}},\ }\bibfield  {title}
  {\enquote {\bibinfo {title} {Spin subdiffusion in the disordered {Hubbard}
  chain},}\ }\href {\doibase 10.1103/PhysRevLett.120.246602} {\bibfield
  {journal} {\bibinfo  {journal} {Phys. Rev. Lett.}\ }\textbf {\bibinfo
  {volume} {120}},\ \bibinfo {pages} {246602} (\bibinfo {year}
  {2018})}\BibitemShut {NoStop}%
\bibitem [{\citenamefont {Prelov\ifmmode~\check{s}\else \v{s}\fi{}ek}\ and\
  \citenamefont {Herbrych}(2017)}]{prelovsek217}%
  \BibitemOpen
  \bibfield  {author} {\bibinfo {author} {\bibfnamefont {P.}~\bibnamefont
  {Prelov\ifmmode~\check{s}\else \v{s}\fi{}ek}}\ and\ \bibinfo {author}
  {\bibfnamefont {J.}~\bibnamefont {Herbrych}},\ }\bibfield  {title} {\enquote
  {\bibinfo {title} {Self-consistent approach to many-body localization and
  subdiffusion},}\ }\href {\doibase 10.1103/PhysRevB.96.035130} {\bibfield
  {journal} {\bibinfo  {journal} {Phys. Rev. B}\ }\textbf {\bibinfo {volume}
  {96}},\ \bibinfo {pages} {035130} (\bibinfo {year} {2017})}\BibitemShut
  {NoStop}%
\bibitem [{\citenamefont {Lev}\ \emph {et~al.}(2017)\citenamefont {Lev},
  \citenamefont {Kennes}, \citenamefont {Kl{\"o}ckner}, \citenamefont
  {Reichman},\ and\ \citenamefont {Karrasch}}]{new_karrasch}%
  \BibitemOpen
  \bibfield  {author} {\bibinfo {author} {\bibfnamefont {Y.~Bar}\ \bibnamefont
  {Lev}}, \bibinfo {author} {\bibfnamefont {D.~M.}\ \bibnamefont {Kennes}},
  \bibinfo {author} {\bibfnamefont {C.}~\bibnamefont {Kl{\"o}ckner}}, \bibinfo
  {author} {\bibfnamefont {D.~R.}\ \bibnamefont {Reichman}}, \ and\ \bibinfo
  {author} {\bibfnamefont {C.}~\bibnamefont {Karrasch}},\ }\bibfield  {title}
  {\enquote {\bibinfo {title} {Transport in quasiperiodic interacting systems:
  From superdiffusion to subdiffusion},}\ }\href
  {http://stacks.iop.org/0295-5075/119/i=3/a=37003} {\bibfield  {journal}
  {\bibinfo  {journal} {EPL (Europhysics Letters)}\ }\textbf {\bibinfo {volume}
  {119}},\ \bibinfo {pages} {37003} (\bibinfo {year} {2017})}\BibitemShut
  {NoStop}%
\bibitem [{\citenamefont {Prelov\ifmmode~\check{s}\else \v{s}\fi{}ek}\ \emph
  {et~al.}(2018)\citenamefont {Prelov\ifmmode~\check{s}\else \v{s}\fi{}ek},
  \citenamefont {Bon\ifmmode~\check{c}\else \v{c}\fi{}a},\ and\ \citenamefont
  {Mierzejewski}}]{prelovsek2018a}%
  \BibitemOpen
  \bibfield  {author} {\bibinfo {author} {\bibfnamefont {P.}~\bibnamefont
  {Prelov\ifmmode~\check{s}\else \v{s}\fi{}ek}}, \bibinfo {author}
  {\bibfnamefont {J.}~\bibnamefont {Bon\ifmmode~\check{c}\else \v{c}\fi{}a}}, \
  and\ \bibinfo {author} {\bibfnamefont {M.}~\bibnamefont {Mierzejewski}},\
  }\bibfield  {title} {\enquote {\bibinfo {title} {Transient and persistent
  particle subdiffusion in a disordered chain coupled to bosons},}\ }\href
  {\doibase 10.1103/PhysRevB.98.125119} {\bibfield  {journal} {\bibinfo
  {journal} {Phys. Rev. B}\ }\textbf {\bibinfo {volume} {98}},\ \bibinfo
  {pages} {125119} (\bibinfo {year} {2018})}\BibitemShut {NoStop}%
\bibitem [{\citenamefont {Agarwal}\ \emph {et~al.}(2016)\citenamefont
  {Agarwal}, \citenamefont {Altman}, \citenamefont {Demler}, \citenamefont
  {Gopalakrishnan}, \citenamefont {Huse},\ and\ \citenamefont
  {Knap}}]{agarwal16}%
  \BibitemOpen
  \bibfield  {author} {\bibinfo {author} {\bibfnamefont {K.}~\bibnamefont
  {Agarwal}}, \bibinfo {author} {\bibfnamefont {E.}~\bibnamefont {Altman}},
  \bibinfo {author} {\bibfnamefont {E.}~\bibnamefont {Demler}}, \bibinfo
  {author} {\bibfnamefont {S.}~\bibnamefont {Gopalakrishnan}}, \bibinfo
  {author} {\bibfnamefont {D.~A.}\ \bibnamefont {Huse}}, \ and\ \bibinfo
  {author} {\bibfnamefont {M.}~\bibnamefont {Knap}},\ }\bibfield  {title}
  {\enquote {\bibinfo {title} {Rare‐region effects and dynamics near the
  many‐body localization transition},}\ }\href {\doibase
  10.1002/andp.201600326} {\bibfield  {journal} {\bibinfo  {journal} {Annalen
  der Physik}\ }\textbf {\bibinfo {volume} {529}},\ \bibinfo {pages} {1600326}
  (\bibinfo {year} {2016})}\BibitemShut {NoStop}%
\bibitem [{\citenamefont {L\"uschen}\ \emph {et~al.}(2017)\citenamefont
  {L\"uschen}, \citenamefont {Bordia}, \citenamefont {Scherg}, \citenamefont
  {Alet}, \citenamefont {Altman}, \citenamefont {Schneider},\ and\
  \citenamefont {Bloch}}]{luschen17}%
  \BibitemOpen
  \bibfield  {author} {\bibinfo {author} {\bibfnamefont {H.~P.}\ \bibnamefont
  {L\"uschen}}, \bibinfo {author} {\bibfnamefont {P.}~\bibnamefont {Bordia}},
  \bibinfo {author} {\bibfnamefont {S.}~\bibnamefont {Scherg}}, \bibinfo
  {author} {\bibfnamefont {F.}~\bibnamefont {Alet}}, \bibinfo {author}
  {\bibfnamefont {E.}~\bibnamefont {Altman}}, \bibinfo {author} {\bibfnamefont
  {U.}~\bibnamefont {Schneider}}, \ and\ \bibinfo {author} {\bibfnamefont
  {I.}~\bibnamefont {Bloch}},\ }\bibfield  {title} {\enquote {\bibinfo {title}
  {Observation of slow dynamics near the many-body localization transition in
  one-dimensional quasiperiodic systems},}\ }\href {\doibase
  10.1103/PhysRevLett.119.260401} {\bibfield  {journal} {\bibinfo  {journal}
  {Phys. Rev. Lett.}\ }\textbf {\bibinfo {volume} {119}},\ \bibinfo {pages}
  {260401} (\bibinfo {year} {2017})}\BibitemShut {NoStop}%
\bibitem [{\citenamefont {De~Roeck}\ and\ \citenamefont
  {Huveneers}(2017)}]{derek}%
  \BibitemOpen
  \bibfield  {author} {\bibinfo {author} {\bibfnamefont {Wojciech}\
  \bibnamefont {De~Roeck}}\ and\ \bibinfo {author} {\bibfnamefont {Fran\ifmmode
  \mbox{\c{c}}\else~\c{c}\fi{}ois}\ \bibnamefont {Huveneers}},\ }\bibfield
  {title} {\enquote {\bibinfo {title} {Stability and instability towards
  delocalization in many-body localization systems},}\ }\href {\doibase
  10.1103/PhysRevB.95.155129} {\bibfield  {journal} {\bibinfo  {journal} {Phys.
  Rev. B}\ }\textbf {\bibinfo {volume} {95}},\ \bibinfo {pages} {155129}
  (\bibinfo {year} {2017})}\BibitemShut {NoStop}%
\bibitem [{\citenamefont {Mierzejewski}\ \emph {et~al.}(2019)\citenamefont
  {Mierzejewski}, \citenamefont {Prelov\ifmmode~\check{s}\else \v{s}\fi{}ek},\
  and\ \citenamefont {Bon\ifmmode~\check{c}\else
  \v{c}\fi{}a}}]{Mierzejewski2019}%
  \BibitemOpen
  \bibfield  {author} {\bibinfo {author} {\bibfnamefont {M.}~\bibnamefont
  {Mierzejewski}}, \bibinfo {author} {\bibfnamefont {P.}~\bibnamefont
  {Prelov\ifmmode~\check{s}\else \v{s}\fi{}ek}}, \ and\ \bibinfo {author}
  {\bibfnamefont {J.}~\bibnamefont {Bon\ifmmode~\check{c}\else \v{c}\fi{}a}},\
  }\bibfield  {title} {\enquote {\bibinfo {title} {Einstein relation for a
  driven disordered quantum chain in the subdiffusive regime},}\ }\href
  {\doibase 10.1103/PhysRevLett.122.206601} {\bibfield  {journal} {\bibinfo
  {journal} {Phys. Rev. Lett.}\ }\textbf {\bibinfo {volume} {122}},\ \bibinfo
  {pages} {206601} (\bibinfo {year} {2019})}\BibitemShut {NoStop}%
\bibitem [{\citenamefont {Bon\ifmmode~\check{c}\else \v{c}\fi{}a}\ \emph
  {et~al.}(2018)\citenamefont {Bon\ifmmode~\check{c}\else \v{c}\fi{}a},
  \citenamefont {Trugman},\ and\ \citenamefont {Mierzejewski}}]{bonca2018}%
  \BibitemOpen
  \bibfield  {author} {\bibinfo {author} {\bibfnamefont {J.}~\bibnamefont
  {Bon\ifmmode~\check{c}\else \v{c}\fi{}a}}, \bibinfo {author} {\bibfnamefont
  {S.~A.}\ \bibnamefont {Trugman}}, \ and\ \bibinfo {author} {\bibfnamefont
  {M.}~\bibnamefont {Mierzejewski}},\ }\bibfield  {title} {\enquote {\bibinfo
  {title} {Dynamics of the one-dimensional {Anderson} insulator coupled to
  various bosonic baths},}\ }\href {\doibase 10.1103/PhysRevB.97.174202}
  {\bibfield  {journal} {\bibinfo  {journal} {Phys. Rev. B}\ }\textbf {\bibinfo
  {volume} {97}},\ \bibinfo {pages} {174202} (\bibinfo {year}
  {2018})}\BibitemShut {NoStop}%
\bibitem [{sup()}]{suppmat}%
  \BibitemOpen
  \href@noop {} {}\bibinfo {note} {See {Supplemental Material} for derivation
  of the transition rates and the details of calculations for disordered
  quantum spin systems}\BibitemShut {NoStop}%
\bibitem [{\citenamefont {Bari\ifmmode \check{s}\else
  \v{s}\fi{}i\ifmmode~\acute{c}\else \'{c}\fi{}}\ and\ \citenamefont
  {Prelov\ifmmode~\check{s}\else \v{s}\fi{}ek}(2010)}]{barisic10}%
  \BibitemOpen
  \bibfield  {author} {\bibinfo {author} {\bibfnamefont {O.~S.}\ \bibnamefont
  {Bari\ifmmode \check{s}\else \v{s}\fi{}i\ifmmode~\acute{c}\else \'{c}\fi{}}}\
  and\ \bibinfo {author} {\bibfnamefont {P.}~\bibnamefont
  {Prelov\ifmmode~\check{s}\else \v{s}\fi{}ek}},\ }\bibfield  {title} {\enquote
  {\bibinfo {title} {Conductivity in a disordered one-dimensional system of
  interacting fermions},}\ }\href {\doibase 10.1103/PhysRevB.82.161106}
  {\bibfield  {journal} {\bibinfo  {journal} {Phys. Rev. B}\ }\textbf {\bibinfo
  {volume} {82}},\ \bibinfo {pages} {161106} (\bibinfo {year}
  {2010})}\BibitemShut {NoStop}%
\bibitem [{\citenamefont {Prelov\ifmmode~\check{s}\else \v{s}\fi{}ek}\ \emph
  {et~al.}(2016{\natexlab{b}})\citenamefont {Prelov\ifmmode~\check{s}\else
  \v{s}\fi{}ek}, \citenamefont {Mierzejewski}, \citenamefont {Bari\ifmmode
  \check{s}\else \v{s}\fi{}i\ifmmode~\acute{c}\else \'{c}\fi{}},\ and\
  \citenamefont {Herbrych}}]{prelovsek116}%
  \BibitemOpen
  \bibfield  {author} {\bibinfo {author} {\bibfnamefont {P.}~\bibnamefont
  {Prelov\ifmmode~\check{s}\else \v{s}\fi{}ek}}, \bibinfo {author}
  {\bibfnamefont {M.}~\bibnamefont {Mierzejewski}}, \bibinfo {author}
  {\bibfnamefont {O.~S.}\ \bibnamefont {Bari\ifmmode \check{s}\else
  \v{s}\fi{}i\ifmmode~\acute{c}\else \'{c}\fi{}}}, \ and\ \bibinfo {author}
  {\bibfnamefont {J.}~\bibnamefont {Herbrych}},\ }\bibfield  {title} {\enquote
  {\bibinfo {title} {Density correlations and transport in models of
  many‐body localization},}\ }\href@noop {} {\bibfield  {journal} {\bibinfo
  {journal} {Annalen der Physik}\ }\textbf {\bibinfo {volume} {529}},\ \bibinfo
  {pages} {1600362} (\bibinfo {year} {2016}{\natexlab{b}})}\BibitemShut
  {NoStop}%
\bibitem [{\citenamefont {{\v S}untajs}\ \emph {et~al.}(2019)\citenamefont {{\v
  S}untajs}, \citenamefont {Bon{\v c}a}, \citenamefont {Prosen},\ and\
  \citenamefont {Vidmar}}]{lev2019}%
  \BibitemOpen
  \bibfield  {author} {\bibinfo {author} {\bibfnamefont {J.}~\bibnamefont {{\v
  S}untajs}}, \bibinfo {author} {\bibfnamefont {J.}~\bibnamefont {Bon{\v c}a}},
  \bibinfo {author} {\bibfnamefont {T.}~\bibnamefont {Prosen}}, \ and\ \bibinfo
  {author} {\bibfnamefont {L.}~\bibnamefont {Vidmar}},\ }\href@noop {}
  {\enquote {\bibinfo {title} {Quantum chaos challenges many-body
  localization},}\ } (\bibinfo {year} {2019}),\ \Eprint
  {http://arxiv.org/abs/1905.06345} {arXiv:1905.06345 [cond-mat.str-el]}
  \BibitemShut {NoStop}%
\bibitem [{\citenamefont {Khemani}\ \emph {et~al.}(2017)\citenamefont
  {Khemani}, \citenamefont {Lim}, \citenamefont {Sheng},\ and\ \citenamefont
  {Huse}}]{Husex2017}%
  \BibitemOpen
  \bibfield  {author} {\bibinfo {author} {\bibfnamefont {Vedika}\ \bibnamefont
  {Khemani}}, \bibinfo {author} {\bibfnamefont {S.~P.}\ \bibnamefont {Lim}},
  \bibinfo {author} {\bibfnamefont {D.~N.}\ \bibnamefont {Sheng}}, \ and\
  \bibinfo {author} {\bibfnamefont {David~A.}\ \bibnamefont {Huse}},\
  }\bibfield  {title} {\enquote {\bibinfo {title} {Critical properties of the
  many-body localization transition},}\ }\href {\doibase
  10.1103/PhysRevX.7.021013} {\bibfield  {journal} {\bibinfo  {journal} {Phys.
  Rev. X}\ }\textbf {\bibinfo {volume} {7}},\ \bibinfo {pages} {021013}
  (\bibinfo {year} {2017})}\BibitemShut {NoStop}%
\bibitem [{\citenamefont {Bouchaud}\ and\ \citenamefont
  {Georges}(1989)}]{bouchaud89}%
  \BibitemOpen
  \bibfield  {author} {\bibinfo {author} {\bibfnamefont {J.~P.}\ \bibnamefont
  {Bouchaud}}\ and\ \bibinfo {author} {\bibfnamefont {A.}~\bibnamefont
  {Georges}},\ }\bibfield  {title} {\enquote {\bibinfo {title} {Anomalous
  diffusion in disordered media: Statistical mechanisms, models and physical
  applications},}\ }\href {\doibase 10.1016/0370-1573(90)90099-N} {\bibfield
  {journal} {\bibinfo  {journal} {Physics Reports}\ }\textbf {\bibinfo {volume}
  {195}},\ \bibinfo {pages} {1} (\bibinfo {year} {1989})}\BibitemShut {NoStop}%
\bibitem [{\citenamefont {\ifmmode~\acute{S}\else \'{S}\fi{}roda}\ \emph
  {et~al.}(2019)\citenamefont {\ifmmode~\acute{S}\else \'{S}\fi{}roda},
  \citenamefont {Prelov\ifmmode~\check{s}\else \v{s}\fi{}ek},\ and\
  \citenamefont {Mierzejewski}}]{sroda19}%
  \BibitemOpen
  \bibfield  {author} {\bibinfo {author} {\bibfnamefont {M.}~\bibnamefont
  {\ifmmode~\acute{S}\else \'{S}\fi{}roda}}, \bibinfo {author} {\bibfnamefont
  {P.}~\bibnamefont {Prelov\ifmmode~\check{s}\else \v{s}\fi{}ek}}, \ and\
  \bibinfo {author} {\bibfnamefont {M.}~\bibnamefont {Mierzejewski}},\
  }\bibfield  {title} {\enquote {\bibinfo {title} {Instability of subdiffusive
  spin dynamics in strongly disordered {Hubbard} chain},}\ }\href {\doibase
  10.1103/PhysRevB.99.121110} {\bibfield  {journal} {\bibinfo  {journal} {Phys.
  Rev. B}\ }\textbf {\bibinfo {volume} {99}},\ \bibinfo {pages} {121110(R)}
  (\bibinfo {year} {2019})}\BibitemShut {NoStop}%
\bibitem [{\citenamefont {Kozarzewski}\ \emph {et~al.}(2019)\citenamefont
  {Kozarzewski}, \citenamefont {Mierzejewski},\ and\ \citenamefont
  {Prelov\ifmmode~\check{s}\else \v{s}\fi{}ek}}]{kozarzewski2019}%
  \BibitemOpen
  \bibfield  {author} {\bibinfo {author} {\bibfnamefont {Maciej}\ \bibnamefont
  {Kozarzewski}}, \bibinfo {author} {\bibfnamefont {Marcin}\ \bibnamefont
  {Mierzejewski}}, \ and\ \bibinfo {author} {\bibfnamefont {Peter}\
  \bibnamefont {Prelov\ifmmode~\check{s}\else \v{s}\fi{}ek}},\ }\bibfield
  {title} {\enquote {\bibinfo {title} {Suppressed energy transport in the
  strongly disordered {Hubbard} chain},}\ }\href {\doibase
  10.1103/PhysRevB.99.241113} {\bibfield  {journal} {\bibinfo  {journal} {Phys.
  Rev. B}\ }\textbf {\bibinfo {volume} {99}},\ \bibinfo {pages} {241113}
  (\bibinfo {year} {2019})}\BibitemShut {NoStop}%
\bibitem [{\citenamefont {Protopopov}\ \emph {et~al.}(2019)\citenamefont
  {Protopopov}, \citenamefont {Panda}, \citenamefont {Parolini}, \citenamefont
  {Scardicchio}, \citenamefont {Demler},\ and\ \citenamefont
  {Abanin}}]{protopopov2019}%
  \BibitemOpen
  \bibfield  {author} {\bibinfo {author} {\bibfnamefont {I.~V.}\ \bibnamefont
  {Protopopov}}, \bibinfo {author} {\bibfnamefont {R.~K.}\ \bibnamefont
  {Panda}}, \bibinfo {author} {\bibfnamefont {T.}~\bibnamefont {Parolini}},
  \bibinfo {author} {\bibfnamefont {A.}~\bibnamefont {Scardicchio}}, \bibinfo
  {author} {\bibfnamefont {E.}~\bibnamefont {Demler}}, \ and\ \bibinfo {author}
  {\bibfnamefont {D.~A.}\ \bibnamefont {Abanin}},\ }\href@noop {} {\enquote
  {\bibinfo {title} {Non-abelian symmetries and disorder: a broad non-ergodic
  regime and anomalous thermalization},}\ } (\bibinfo {year} {2019}),\ \Eprint
  {http://arxiv.org/abs/1902.09236} {arXiv:1902.09236 [cond-mat.str-el]}
  \BibitemShut {NoStop}%
\bibitem [{\citenamefont {Imbrie}(2016{\natexlab{b}})}]{imbrie14}%
  \BibitemOpen
  \bibfield  {author} {\bibinfo {author} {\bibfnamefont {John~Z.}\ \bibnamefont
  {Imbrie}},\ }\bibfield  {title} {\enquote {\bibinfo {title} {On many-body
  localization for quantum spin chains},}\ }\href {\doibase
  10.1007/s10955-016-1508-x} {\bibfield  {journal} {\bibinfo  {journal}
  {Journal of Statistical Physics}\ }\textbf {\bibinfo {volume} {163}},\
  \bibinfo {pages} {998--1048} (\bibinfo {year}
  {2016}{\natexlab{b}})}\BibitemShut {NoStop}%
\bibitem [{\citenamefont {Prelov\v{s}ek}\ and\ \citenamefont
  {Bon\v{c}a}(2013)}]{prelovsek13}%
  \BibitemOpen
  \bibfield  {author} {\bibinfo {author} {\bibfnamefont {P.}~\bibnamefont
  {Prelov\v{s}ek}}\ and\ \bibinfo {author} {\bibfnamefont {J.}~\bibnamefont
  {Bon\v{c}a}},\ }\bibfield  {title} {\enquote {\bibinfo {title} {Ground state
  and finite temperature lanczos methods},}\ }\bibfield  {booktitle} {\emph
  {\bibinfo {booktitle} {Strongly Correlated Systems - Numerical Methods}},\
  }\href@noop {} {\  (\bibinfo {year} {2013})}\BibitemShut {NoStop}%
\end{thebibliography}%
\clearpage
\newpage
\setcounter{figure}{0}
\setcounter{equation}{0}
\setcounter{page}{1}

\renewcommand{\thetable}{S\arabic{table}}
\renewcommand{\thefigure}{S\arabic{figure}}
\renewcommand{\theequation}{S\arabic{equation}}
\renewcommand{\thepage}{S\arabic{page}}

\renewcommand{\thesection}{S\arabic{section}}

\onecolumngrid

\begin{center}

{\large \bf Supplemental Material,\\
Resistivity and its fluctuations in disordered many-body systems: \\ from chains to planes}\\

\vspace{0.3cm}

{M. Mierzejewski$^{1}$, M. {\'S}roda$^{1}$, J. Herbrych$^{1}$, P. Prelov\v sek$^{2,3}$} \\
\vspace{0.2cm}
$^1${\it Department of Theoretical Physics, Faculty of Fundamental Problems of Technology, \\ Wroc\l aw University of Science and Technology, 50-370 Wroc\l aw, Poland}\\
$^2${\it Department of Theoretical Physics, J. Stefan Institute, SI-1000 Ljubljana, Slovenia} \\
$^3${\it Department of Physics, Faculty of Mathematics and Physics, University of Ljubljana, SI-1000 Ljubljana, Slovenia}

\end{center}

\vspace{0.6cm}

\twocolumngrid

\label{pagesupp}

\section{Transition Rates for single particle coupled to hard-core bosons} \label{app1}

We recall the main steps of derivations in \cite{prelovsek2018a} and \cite{Mierzejewski2019} for the transition rates between the Anderson states originating from the coupling to hard-core bosons. To this end, we solve the single-particle eigenproblem 
\begin{equation} 
 - t \sum_{\langle i,j \rangle} c^\dagger_{i} c^{\phantom{\dagger}}_j + \sum_j \varepsilon_j n_j =\sum_{l} \epsilon_l \varphi^{\dagger}_l \varphi_l^{\phantom{\dagger}},
\end{equation}
where $ \varphi_l^\dagger = \sum_i \phi_{li} c_i^\dagger$ creates a particle in the Anderson-localized state with the eigenfunction $\phi_{li}$ and we take all $\phi_{li}$ as real. Then, we rewrite the interaction part of the Hamiltonian using $ \varphi_l^\dagger$ 
\begin{eqnarray}
H' &=& \sum_{i,l,k} \eta_{l k i} \varphi_{l}^\dagger \varphi_{k}^{\phantom{\dagger}} (a^\dagger_i + a_i^{\phantom{\dagger}}) , \quad \quad \eta_{l k i} = g \phi_{ki} \phi_{li}. \label{smham1}
\end{eqnarray}
We use the Fermi golden rule (FGR) to calculate the transition rate from $|l\rangle$ to $|k\rangle$
 \begin{eqnarray}
\Gamma_{l k} & = & 2 \pi \sum_{b,a} w_b | \langle l,b | H' | k,a \rangle |^2 \delta(E_{b,l}-E_{a,k}), \nonumber \\
=&& \sum_{b} w_b \int_{-\infty}^{\infty} {\rm d} t\,\langle l,b | H'(t) \left( | k \rangle \langle k |\otimes {\cal I}_b \right) H' | l,b \rangle, \nonumber \\
\label{smfgr}
\end{eqnarray}
where $| l,b \rangle= | l \rangle \otimes |b \rangle $ and $w_b$ are the equilibrium probabilities for finding the hard-core bosons in the state $ |b \rangle $. Here, we consider only the case of infinite temperature $T \rightarrow \infty$, hence $w_b=\text{const}$. 

For hard-core bosons, there may be at most a single boson creation/annihilation per site. Therefore, multi-boson contributions to FGR are significantly reduced with respect to regular bosons. This reduction is particularly important for strong disorder, i.e., for a short localization length of the Anderson states $\phi_{li}$. Neglecting the multi-boson contributions, we rewrite the perturbation using the wave-vector representation for the bosonic operators
\begin{eqnarray}
H'(t) & \simeq & \sum_{l k \bf{q}} \eta_{lk\bf{q}} \varphi_{l}^\dagger \varphi_k^{\phantom{\dagger}} [ a_{-\bf{q}}^{\phantom{\dagger}} e^{it(\epsilon_l- \epsilon_{k} -\omega_{\bf{q}})}+a^{\dagger}_{\bf{q}} e^{it(\epsilon_l- \epsilon_{k}+\omega_{\bf{q}})} ], \nonumber \\
\label{smham2}
\end{eqnarray} 
where 
\begin{eqnarray}
\omega_{\bf{q}} & = & \omega_0-2 t_b[\cos(q_x)+\cos(q_y)] , \\
 \eta_{lk\bf{q}} & = &\frac{g}{\sqrt{N}} \sum_j e^{ - i \bf{q}\cdot \bf{R}_j } \phi_{lj } \phi_{kj },
\end{eqnarray}
c.f. Hamiltonian (\ref{ham}) in the main text. At $T\rightarrow \infty$ the hard-core bosons randomly occupy the single-particle states, thus
\begin{eqnarray}
\sum_b w_b \langle b | a^{\dagger}_{\bf{q}} a^{\phantom{\dagger}}_{\bf{q'}} | b \rangle = \sum_b w_b \langle b | a^{\phantom{\dagger}}_{\bf{q}} a^{\dagger}_{\bf{q'}} | b \rangle=\frac{1}{2} \delta_{\bf{q},\bf{q'}}
\label{smoccup}
\end{eqnarray} 
Substituting Eq.~(\ref{smham2}) into (\ref{smfgr}) and using (\ref{smoccup}) one finds the transition rates 
\begin{eqnarray}
\Gamma_{l k} & = & \pi \sum_{\bf{q}} | \eta_{lk\bf{q}} |^2 \left[ \delta(\epsilon_l- \epsilon_{k} -\omega_{\bf{q}}) + \delta(\epsilon_l- \epsilon_{k}+\omega_{\bf{q}}) \right], \nonumber \\
\label{smfgr1}
\end{eqnarray}
which take into account the bosonic dispersion relation and the details of the matrix elements $ \eta_{l k \bf{q}} $. In the main text, we refer to Eq.~(\ref{smfgr1}) as FGR. Using FGR we study systems up to $N \sim 10^3$.

One may significantly simplify the numerical calculations by neglecting the $\bf{q}$-dependence of the matrix elements
\begin{equation}
| \eta_{lk\bf{q}} |^2 \simeq | \eta_{lk}|^2=\frac{1}{N} \sum_{\bf{q}} | \eta_{lk\bf{q}} |^2=\frac{g^2}{N} \sum_j (\phi_{lj } \phi_{kj })^2,
\end{equation}
and assuming a uniform bosonic density of states
\begin{equation}
\frac{1}{N} \sum_q \delta(\omega-\omega_q) \simeq \frac{1}{\Omega} \theta(\Omega-\omega), \quad \quad \omega \ge 0,
\end{equation}
where $\Omega$ is an effective bosonic frequency. Within these simplifcations the transition rates read
\begin{eqnarray}
\Gamma_{l k} & = & \pi \frac{g^2}{\Omega} \theta(\Omega-| \epsilon_l- \epsilon_{k}| ) \sum_j (\phi_{lj } \phi_{kj })^2.
\label{smfgr2}
\end{eqnarray}
We refer to Eq.~(\ref{smfgr2}) as the simplified FGR (SFGR) for which we are able to reach $N \sim 10^4$. In the numerical calculations we take $\omega_0=g=1$ and $t_b=0.2$. In the quasi-2D system, the bosonic spectrum has width $8t_b$ and the exact density of states is strongly peaked at $\omega=\omega_0$. For this reason we take $\Omega=1.2 < 8t_b$. 

In order to find the stationary solution of RE~(\ref{req}) in the main text, we put $\frac{{\rm d} n_l}{{\rm d} t}=0$ and diagonalize 
the matrix 
\begin{eqnarray}
\widetilde{\Gamma}_{lk} &=& \delta_{lk}\sum_{j} \Gamma_{lj}- \Gamma_{kl}, \\
 \widehat{\widetilde{\Gamma}}&=& \widehat{U} \; {\rm diag}(\lambda_1,...\;,\lambda_N) \; \widehat{U}^T. 
 \end{eqnarray}  
Then, one obtains the stationary occupations of the  Anderson states
\begin{equation}
n_l=\sum_{l,k,j} U_{l j}\frac{1}{\lambda_j}U^{T}_{j k}I_{k}, 
\end{equation}
where we have omitted the zero-mode, $\lambda_1=0$, corresponding to the conservation of the total particle number. In order to eliminate the boundary effects, we divide the system into three sections of equal size. The gradient of the particle density  in real space, $\nabla n_i$, is obtained from the linear fitting of $n_i=\sum_l n_l |\phi_{li}|^2$ in the middle section.

\section{Statistical fluctuations of conductivity in disordered spin chains} \label{app2}

\begin{figure*}[!htb]
 \includegraphics[width=\textwidth]{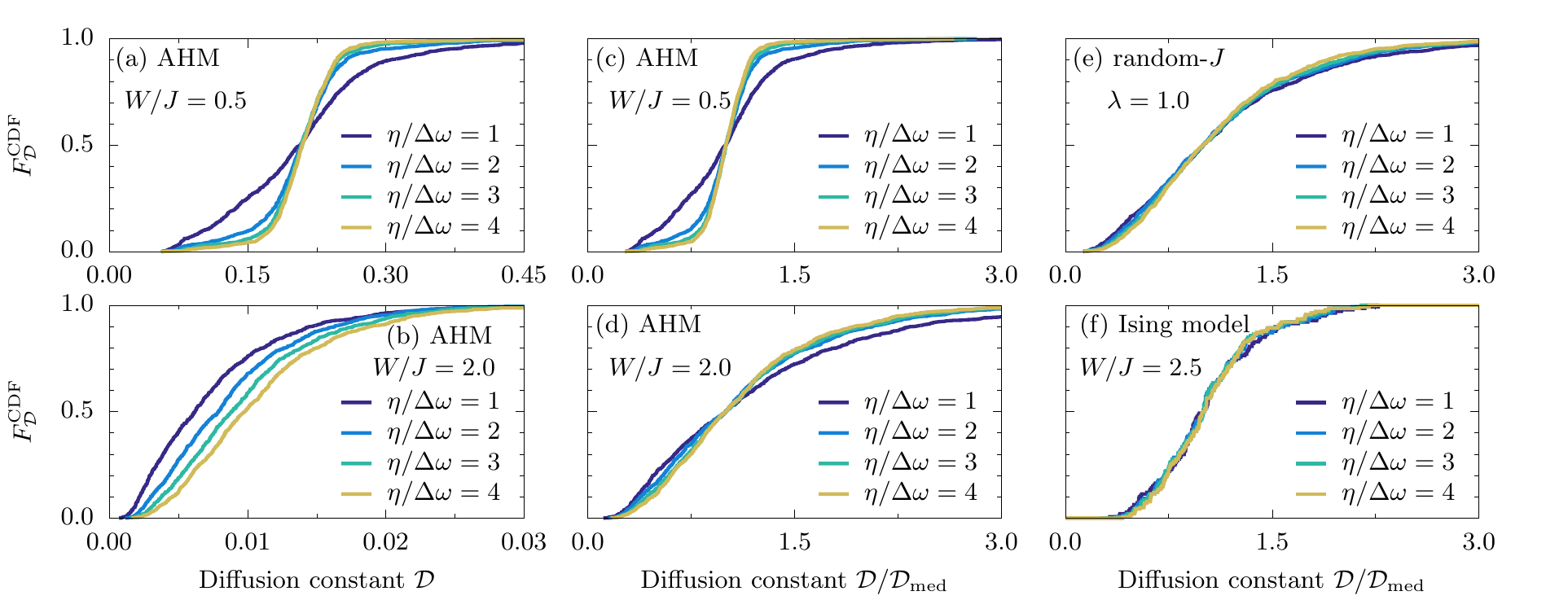}
\caption{Cumulative distribution functions for the diffusion constant, $F_{\cal D}({\cal D})$, as calculated for various disorder strengths and broadenings $\eta$ for (a-d) random-field Heisenberg model, (e) ferromagnetic random-$J$ Heisenberg model (effective spin Hamiltonian of the Hubbard model with strong disorder in the charge potential), and (f) random transverse-field Ising model (see the text for details). }
\label{figS}
\end{figure*} 

In the main text of the manuscript, we have considered three one-dimensional quantum spin chains with quenched disorder:

\noindent {\bf (i)} Antiferromagnetic Heisenberg model (AHM) 
\begin{eqnarray}
H&=& J\sum_{i}\left(S^{x}_{i}S^{x}_{i+1} + S^{y}_{i}S^{y}_{i+1} + \Delta S^{z}_{i}S^{z}_{i+1}\right)\nonumber\\
 &+& J\Delta_2 \sum_{i} S^{z}_{i}S^{z}_{i+2}+\sum_{i} h_{i}\,S^{z}_{i}\,,
\end{eqnarray}
where we have set $J=1$ as the unit of energy and the local magnetic fields $h_{i}$ are random numbers drawn from a uniform distribution in the interval $h_{i}\in [-W,W]$. Furthermore, we have chosen $\Delta=0.75$ and $\Delta_2=0.5$. The latter is the integrability breaking term in the clean limit, $W=0$.

\noindent {\bf (ii)} Ferromagnetic random-$J$ Heisenberg model (effective spin Hamiltonian of the Hubbard model with strong disorder in the charge potential, see Ref.~[\onlinecite{kozarzewski18}] and [\onlinecite{sroda19}] of the main text for more details), i.e.,
\begin{eqnarray}
H&=&-\sum_{i} J_{i}\,\mathbf{S}_{i}\cdot\mathbf{S}_{i+1}\,.
\end{eqnarray}
Here $\mathbf{S}_i$ stands for $\mathbf{S}_{i}=(S^x_i,S^y_i,S^z_i)$ and exchange coupling $J_{i}$ has to be drawn from the probability distribution given by \mbox{$f_J(J)=\lambda J^{\lambda-1}$} for $0\leqslant J\leqslant 1$, where $\lambda$ controls the disorder strength (we refer the interested reader to Ref.~[\onlinecite{kozarzewski18}] and [\onlinecite{sroda19}] of the main text for details on this model).

\noindent {\bf (iii)} Random transverse-field Ising model
\begin{equation}
H = \sum_{i} (J + \delta J_i) S^{z}_{i} S^{z}_{i+1} + \sum_{i} h_{i}\,S^{z}_{i} +   f \sum_{i} \,S^{x}_{i}\,,
\end{equation}
with $J=1$ as the unit of energy, fixed parameter $f/J =0.5$, and uniform distribution $h_{i}\in [-W,W]$. Similarly to the AHM we break the integrability of the clean case ($W=0$), i.e.,  we added small randomness in the spin exchange coupling, $\delta  J_{i}\in [-W_J,W_J]$ with $W_J/J=0.2$, thus varying only $W$ in our consideration.

The regular part of the generic conductivity $C(\omega)$ in the high-temperature limit ($\beta\to 0$) can be defined as follows:
\begin{equation}
C_{\mathrm{reg}}(\omega)=\frac{\pi}{LZ}\sum_{E_n\ne E_m} |\langle n|j|m\rangle|^2\,\delta(E_n-E_m-\omega)\,,
\label{scond}
\end{equation}
where $L$ is the considered system size [$L=26$ for the models (i) and (ii); $L= 20$ for the model (iii)], $Z$ is the partition function (for $\beta\to 0$, $Z$ is the dimension of the Hilbert space), $|n\rangle$ and $E_n$ are the many-body eigenstates and the eigenvalues, respectively. For the Heisenberg models [(i) and (ii)] the diffusion constant was extracted from the spin conductivity \mbox{${\cal D}=C_{\mathrm{reg}}(\omega\to0)=\sigma_{\mathrm{reg}}(\omega\to0)/\beta$} with the spin current operator defined as $j=j^S\equiv \sum_{i}J_{i} \left(S^{x}_{i}S^{y}_{i+1}-S^{y}_{i}S^{x}_{i+1}\right)$. For the transverse-field Ising model (iii) the only conserved quantity is energy. As a consequence, we evaluate the enery (thermal) diffusion constant obtained from the thermal conductivity \mbox{${\cal D}=C_{\mathrm{reg}}(\omega\to0)=\kappa_{\mathrm{reg}}(\omega\to0)/\beta^2$} with the energy current operator defined as $j=j^E\equiv f \sum_{i} (J + \delta J_i)S^z_i {S}^x_{i+1}$. Eq.~\eqref{scond} is then numerically evaluated with the help of the Microcanonical Lanczos Method \cite{prelovsek13} with $M_\mathrm{Lan}=10^4$ Lanczos steps. The latter allows us to obtain frequency resolution \mbox{$\Delta\omega=\Delta E /M_\mathrm{Lan}\simeq 10^{-3}$}, where $\Delta E$ is the energy span. 

It is important to note that the spectrum in Eq.~\eqref{scond} is discrete and in order to properly resolve the $\omega\to 0$ limit one has to artificially broaden it with, e.g., the Gaussian kernel,
\begin{equation}
C^{\mathrm{S}}_{\mathrm{reg}}(\omega)=\int\limits_{-\infty}^{\infty}\mathrm{d}\omega^{\prime} \frac{1}{\sqrt{2\pi}\eta} \mathrm{e}^{-\frac{(\omega-\omega^\prime)^2}{2\eta^2}}  \,C^{\mathrm{R}}_{\mathrm{reg}}(\omega')\,,
\end{equation} 
where $C^{\mathrm{R}}_{\mathrm{reg}}$ ($C^{\mathrm{S}}_{\mathrm{reg}}$) refers to the raw (smoothed) data. As a consequence, the results---especially in the $\omega\to 0$ limit---can be influenced by the broadening $\eta$. In Fig.~\ref{figS} we present the cumulative distribution functions of the diffusion constant, $F_{\cal D}({\cal D})$, for all considered models and various values of the broadening $\eta$. It is evident from the presented results that for large disorder, the distribution $F_{\cal D}({\cal D})$ does indeed depend on the value of $\eta$ [see panels (a) and (b)]. However, our results also indicate [panels (c) and (d)] that for any realistic broadening, $\eta>\Delta\omega$, the normalized by median distribution is $\eta$-independent (i.e., the functional form of $F_{\cal D}({\cal D})$ is almost $\eta$-independent) \cite{barisic16}. Such behavior was observed for all considered models: see panel (e) for the random-$J$ model and panel (f) for the random transverse-field Ising model.

\end{document}